\documentclass[twocolumn]{aastex63}
\usepackage{amsmath}
\usepackage{multirow}
\hypersetup{linkcolor=red,citecolor=magenta,filecolor=cyan,urlcolor=magenta}
\submitjournal{AJ}

\shorttitle{Planets in clusters}
\shortauthors{Yuan-Zhe Dai et al.}
\graphicspath{{./}{figures/}}
\begin{document}

\title{Understanding the Planetary Formation and Evolution in Star Clusters(UPiC)-I: Evidence of Hot Giant Exoplanets Formation Timescales} 

\correspondingauthor{Hui-Gen Liu}
\email{huigen@nju.edu.cn}

\author{Yuan-Zhe Dai}
\affiliation{School of Astronomy and Space Science, Nanjing University, 163 Xianlin Avenue, Nanjing, 210023, People's Republic of China}
\affiliation{Key Laboratory of Modern Astronomy and Astrophysics, Ministry of Education, Nanjing, 210023, People's Republic of China}
\author{Hui-Gen Liu}
\affiliation{School of Astronomy and Space Science, Nanjing University, 163 Xianlin Avenue, Nanjing, 210023, People's Republic of China}
\affiliation{Key Laboratory of Modern Astronomy and Astrophysics, Ministry of Education, Nanjing, 210023, People's Republic of China}
\author{Jia-Yi Yang}
\affiliation{School of Astronomy and Space Science, Nanjing University, 163 Xianlin Avenue, Nanjing, 210023, People's Republic of China}
\affiliation{Key Laboratory of Modern Astronomy and Astrophysics, Ministry of Education, Nanjing, 210023, People's Republic of China}
\author{Ji-Lin Zhou}
\affiliation{School of Astronomy and Space Science, Nanjing University, 163 Xianlin Avenue, Nanjing, 210023, People's Republic of China}
\affiliation{Key Laboratory of Modern Astronomy and Astrophysics, Ministry of Education, Nanjing, 210023, People's Republic of China}

\begin{abstract}
Planets in young star clusters could shed light on planet formation and evolution since star clusters can provide accurate age estimation. However, the number of transiting planets detected in clusters was only $\sim 30$, too small for statistical analysis. Thanks to the unprecedented high-precision astrometric data provided by Gaia DR2 and Gaia DR3, many new Open Clusters(OCs) and comoving groups have been identified. The UPiC project aims to find observational evidence and interpret how planets form and evolve in cluster environments. In this work, we cross-match the stellar catalogs of new OCs and comoving groups with confirmed planets and candidates. We carefully remove false positives and obtain the biggest catalog of planets in star clusters up to now, which consists of 73 confirmed planets and 84 planet candidates. After age validation, we obtain the radius--age diagram of these planets/candidates. We find an increment of the fraction of Hot Jupiters(HJs) around 100 Myr and attribute the increment to the flyby-induced high-e migration in star clusters. An additional small bump of the fraction of HJs after 1 Gyr is detected, which indicates the formation timescale of HJ around field stars is much larger than that in star clusters. Thus, stellar environments play important roles in the formation of HJs. The Hot-Neptune desert occurs around 100 Myr in our sample. Combining photoevaporation and high-e migration may sculpt the Hot-Neptune desert in clusters.

\end{abstract}

%% We recommend that authors also use the natbib \citepp
%% and \citept commands to identify citations.  The citations are

%* []介绍年轻恒星周围行星的进展情况，若干项目
%* []介绍Gaia的数据下星团成员星数目的增加
%* []引出本工作的motivation，more stars in clusters，more planets in clusters or around young stars. 
%* []最后的对本文的结构做一个简单的梳理。

\section{Introduction}
 Open Clusters(OCs) in the Milky Way are the collection of stars formed from the same molecular cloud and gravitationally bound together, Thus, sharing similar specific characteristics, e.g. age, distance, reddening, metal abundance, etc. OCs provide an ideal laboratory for studying star formation and evolution. Recent studies based on Kepler data show that nearly 50\% of stars host planets \citep{2021ARA&A..59..291Z}. Since most stars form in clusters \citep{2003ARA&A..41...57L}, many exoplanets are formed in cluster environments. Then the majority of stars will eventually become field stars as clusters are dissociated. Detecting exoplanets in OCs can provide an ideal sample for studying planet formation and evolution.

%ZEIT 9,10,11 not new planet detection
The first planet in OCs, $\epsilon$ tau b, was detected by \cite{2007ApJ...661..527S} via Radial Velocity. Kepler-66b and Kepler-67b are the first cluster planets discovered by transit \citep{2013Natur.499...55M}. Thanks to Kepler/K2 and TESS, tens of planets in clusters have been discovered, and the number is growing. There are several programs focusing on planets in star clusters, especially young exoplanets. Zodiacal Exoplanets In Time (ZEIT) collaboration uses K2 data to monitor young open clusters and associations in the ecliptic plane and found planets in the Hyades, Praesepe, Upper Sco, and Taurus (\citealt{2016ApJ...818...46M,2016AJ....152...61M,2017AJ....153...64M,2018AJ....155....4M,2017AJ....154..224R,2017MNRAS.464..850G,2018AJ....156...46V,2018AJ....156..195R}). With the help of enable extensive follow-up observations, The TESS Hunt for Young and Maturing Exoplanets (THYME) collaboration has reported on planets in Upper Sco \citep{2020AJ....160...33R}, the Tuc-Hor association \citep{2019ApJ...880L..17N}, the Ursa Major moving group \citep{2020AJ....160..179M}, and the Pisces Eridanus stream \citep{2021AJ....161...65N}. \cite{2019ApJS..245...13B,2020AJ....160..239B}  begun a Cluster Difference Imaging Photometric Survey(CDIPS) to  discover giant transiting planets with known ages, and to provide light curves suitable for studies in stellar astrophysics. \citealt{2019MNRAS.490.3806N,2020MNRAS.495.4924N,2020MNRAS.498.5972N,2021MNRAS.505.3767N} use a PSF-based Approach to TESS High-quality data of star clusters (PATHOS) and find 90 planet candidates. \textbf{The GAPS Young Objects project aims to search and characterize young hot Jupiters and put constraints on evolutionary models (e.g.  \cite{2020A&A...638A...5C}). For planets at larger separations, direct imaging plays an important role. There are dozens of young planets discovered through direct imaging e.g. 2MASS J12073346-3932539 b \cite{2004A&A...425L..29C}, DH Tau b \cite{2005ApJ...620..984I}, GQ Lup b \cite{2005A&A...435L..13N}, and etc.}

Hitherto, there are many surveys and projects focusing on the young planets in clusters, but the number of reported planets in clusters is limited and is not enough to support statistical work, $\sim 30$ according to \cite{2021MNRAS.505.3767N}. \textbf{To enlarge the number of planets in open clusters, both expanding the number of stars in clusters and identifying new open clusters are feasible in the Gaia Era.}

%确认是否是1.3billion?已确认
The Gaia DR2/EDR3 catalog \citep{2018A&A...616A...1G,2021A&A...649A...2L} presents more than 1.3 billion stars with unprecedented high-precision astrometric and photometric data, greatly improving the reliability of stellar membership determination and characterization of a large sample of stellar groups including star clusters, association, and other comoving groups. The recent analysis of Gaia Data has greatly expanded our knowledge of stellar groups \citep[e.g.,][]{2018A&A...618A..93C,2019AJ....158..122K,2021ApJ...917...23K}. In previous knowledge, a star cluster is a set of stars that are gravitationally bound to one another \citep{2010ARA&A..48..431P}. However, the recent discovery of stars in diffuse regions reminds us that we need to extend the original definition of star clusters. Because these stars in diffuse regions, i.e. not gravitational bound, are proved to have the same age as core cluster members through analyses of color-absolute magnitude diagrams \citep[CAMDs;][]{2019AJ....158..122K,2019A&A...621L...3M,2021AJ....162..197B}. On top of that, these stars in diffuse regions also share a similar distribution with core cluster members in stellar rotation periods \citep{2021AJ....162..197B} and chemical abundances \citep{2020A&A...635L..13A,2020MNRAS.496.2422H}. Therefore, these stars in diffuse regions are probably coeval. In this series of papers, we extend the definition of open clusters to those stars in diffuse regions, i.e. comoving groups, diffuse streams, tidal tails, etc., not only the core cluster members. Thus, the number of stars in open clusters can be extremely enlarged. 

%To briefly describe stars in these comoving groups, we call them as stars in comoving groups. 

%There are many names to describe such diffuse structures, like diffuse streams and tidal tails are comparable in stellar mass to the previously known cores of nearby open clusters(\cite{2019A&A...622L..13M,2021A&A...645A..84M,2021ApJ...915L..29G}). Even though these streams are spread over tens to hundreds of parsecs, their velocity dispersions can remain coherent at the $\sim$1 km/s$^{-1}$ level. They found many stars in diffuse regions can be verified to be the same age as the core cluster members through analyses of color–absolute magnitude diagrams(CAMDs; \cite{2019AJ....158..122K}), stellar rotation periods(\cite{2021AJ....162..197B}), and chemical abundances(\cite{2002AJ....123..905A,2020MNRAS.496.2422H}). 

% method part
There are many previous works that identified new OCs in the Milky Way using different algorithms, e.g. \cite{2018A&A...618A..93C} applied the UPMASK algorithm to select cluster members and provided an updated catalog of 1229 OCs; \cite{2020A&A...633A..99C} found 582 new cluster candidates located in the low galactic latitude area using an algorithm named Density-Based Spatial Clustering of Applications with Noise(DBSCAN), and etc. Using Hierarchical Density-Based Spatial Clustering of Applications with Noise \citep[HDBSCAN,][]{McInnes2017hdbscanHD}, \cite{2019AJ....158..122K} systematically clustered Gaia DR2 data within 1 kpc and identified 1640 populations containing a total of 288,370 stars. In their recent work, \cite{2020AJ....160..279K} (hereafter K2020), they extended the distance from 1 to 3 kpc and identified 8292 comoving groups consisting of 987,376 stars.

%% move to part 2 ??
%If we combine all the catalogs of stellar groups, definitely, we can get more significant stellar numbers. But the difference in data selection method and clustering algorithms will increase the inhomogeneity of the final catalog. Therefore, in order to maximize the stellar numbers and make the sample as homogeneous as possible, we select K2020, the largest catalog. Then, we use the star catalog in K2020 to cross-match with confirmed planets and planet candidates.

% to strengthen the motivation of UPiC projects in the 1st paper.
%In the Gaia era, the number of stellar groups is largely enhanced. 
Utilizing the enlarged stellar population in clusters, the UPiC (Understanding Planetary Formation and Evolution in Star Clusters) project focuses on the planets in open clusters, including association and other comoving groups. We are trying to find evidence that how planets form and evolve in cluster environments. Lots of works have shown that both dynamical \citep{2009ApJ...697..458S,2013ApJ...772..142L,2017MNRAS.470.4337C,2017AJ....154..272H,2021ApJ...913..104R,2023MNRAS.518.4265L} and radiation \citep{1998ApJ...499..758J,2003ApJ...582..893M,2018MNRAS.480.4080D,2018MNRAS.478.2700W} environments in clusters can influence the planet's formation and evolution. As the initial work of UPiC, this paper collects the largest transiting planet sample in open clusters and aims to analyze the correlation between the planetary radius and cluster ages, which is crucial for planet formation timescales. For example, \citep{2011ApJ...727L..44S,2013ApJ...763...12B,2016A&A...589A..75M} discovered the hot Neptune desert in planetary mass-period and radius-period distribution. The boundaries of the desert can be explained by photoevaporation and high-e migration \cite{2018MNRAS.479.5012O}. These two mechanisms have different timescales. Thus, the age of the planets can distinguish different mechanisms and help us understand the time evolution of planet radius.

This paper is arranged as follows. In section \ref{sec:2}, we describe methods, including data collection, age validation, and sample cut. In section \ref{sec:3}, we use these refined data to study planet radius -- age diagram and estimate the evolution of three different planet populations. In section \ref{sec:4}, we discuss how the statistical results constrain the planet formation mechanisms, including high-e migration and photoevaporation. In section \ref{sec:5}, we discuss some additional influences and caveats. Last, we summarize our major conclusions in section \ref{sec:6}. 

\section{Catalog of Transiting Planet in OCs} \label{sec:2}
\subsection{Data collection}
There are many works that use Gaia data to identify new OCs. If we combine all the catalogs of star clusters, we can definitely get more OCs, so as the stars in OCs. However, data selection criteria and clustering algorithms vary in different works, which will increase the inhomogeneity of the combined catalog. Therefore, to maximize the number of stars in OCs and make the sample as homogeneous as possible, we adopt the catalog from K2020, the largest catalog of stars with age estimations in comoving groups. 
%difference in data selection method and clustering algorithms will increase the inhomogeneity of the final catalog. Therefore, in order to maximize the stellar numbers and make the sample as homogeneous as possible, we select K2020, the largest catalog. Then, we use the star catalog in K2020 to cross-match with confirmed planets and planet candidates.

K2020 identified 8292 comoving groups within 3 kpc and galactic latitude $|b|< 30 ^{\circ}$ by applying the unsupervised machine learning algorithm HDBSCAN on Gaia DR2's 5D data. We use the stellar catalog of K2020 to cross-match with the host stars of confirmed transiting planets and planet candidates. In this section, we use planets/candidates from Kepler, K2, and TESS. Since we are concerned about the radius of planets, we do not consider planets detected via the Radial Velocity method. The following subsections will briefly introduce how we select the planets and cut the planet sample to exclude some observation biases.

%?修改表格
\begin{deluxetable*}{cccccccccccc}
\centering
\tabletypesize{\scriptsize}
\tablewidth{0pt} 
\tablenum{1}
\tablecaption{Planets in clusters \label{tab:pcl}}
\tablehead{
\colhead{Plname}     &\colhead{$R_{\rm p}$}      & \colhead{Period} & \colhead{OName} & \colhead{Group}    & \colhead{Gaia DR2}             & \colhead{Age}  & \colhead{Validation}                 & \colhead{$T_{\rm eff}$} & \colhead{logg} & \colhead{Stmass}  & \colhead{Flag}\\
\colhead{}   &\colhead{($\rm R_{\oplus}$)} & \colhead{(days)}   & \colhead{}      & \colhead{}         &  \colhead{}  & \colhead{(Myr)}                  &  \colhead{} & \colhead{(K)}    & \colhead{}     & \colhead{(M$_{\odot}$)} & \colhead{}
}
\colnumbers
\startdata 
TOI 520.01 &1.49$^{+0.66}_{-0.66}$   & 0.524  & \nodata  & Group 95   & 5576476552334683520 & 30.2$^{+5.3}_{-4.5}$   & NO &7450 & 4.34 & 1.66   &  TESS\\
TOI 626.01 &19.74$^{+0.66}_{-0.66}$  & 4.40   & \nodata  & Group 449  & 5617241426979996800 & 195$^{+80}_{-57}$      & NO &8489 & 4.03 & 2.11   &  TESS\\
TOI 2453.01 &3.02$^{+0.20}_{-0.20}$  & 4.44   & Hyades   & Group 1004 & 3295485490907597696  & 646$^{+113}_{-96}$    & NO &3609 & 4.73 & 0.50   &  TESS\\
TOI 2519.01 &2.29$^{+0.20}_{-0.20}$  & 6.96   & Columba  & Group 208  & 2924619634745251712 & 263$^{+109}_{-77}$     & NO &4742 & 4.57 & 0.76   &  TESS\\
TOI 2640.01 &7.39$^{+0.50}_{-0.50}$  & 0.911  & IC 2602  & Group 92   & 5404579488593432576 & 45$^{+14}_{-11}$       & NO &2999 & 4.95 & 0.25   &  TESS\\
TOI 2646.01 &8.10                    & 0.313  & NGC 2516 & Group 613  & 5288535107223500928 & 145$^{+37}_{-30}$      & NO &5202 & 4.53 & 0.88   &  TESS\\
TOI 2822.01 &11.53$^{+0.72}_{-0.72}$ & 2.88   & \nodata  & Group 5076 & 5597777288033556480 & 537$^{+94}_{-80}$      & NO &6086 & 4.04 & 1.14   &  TESS\\
TOI 3077.01 &13.67$^{+0.72}_{-0.72}$ & 6.36   & \nodata  & Group 3176 & 5307513536932390272 & 186$^{+130}_{-77}$     & NO &7689 & 4.06 & 1.80   &  TESS\\
TOI 3335.01 &11.60$^{+0.70}_{-0.70}$ & 3.61   & \nodata  & Group 550  & 5903623451060661504 & 214$^{+75}_{-55}$      & NO &6071 & 4.04 & 1.13   &  TESS\\
TOI 1097.02 &12.7$^{+1.0}_{-1.0}$    & 2.27   & \nodata  & Group 1502 & 6637496339607744768 & 3020$^{+447}_{-390}$   & NO &5568 & 3.89 & 0.98   &  TESS\\
\nodata &\nodata    & \nodata   & \nodata  & \nodata & \nodata & \nodata   & \nodata &\nodata & \nodata & \nodata   &  \nodata \\
\enddata
\tablecomments{Due to the space limitation, we only show the first 10 rows of the table. Here, ``OName'' shows the name of the parental cluster, ``Group'' shows the corresponding group number in K2020, ``Flag'' shows the name of the mission that detects the planets, and ``Validation'' shows whether this planet/candidate has the convincing age estimation (i.e. ``YES'', ``NO'', and ``Excluded''). The machine-readable table can be obtained in the following link: \href{https://github.com/astrodyz/UPiC}{https://github.com/astrodyz/UPiC}}
\end{deluxetable*}

\subsubsection{K2020 cross-matching with confirmed exoplanets} \label{subsec:2.1.1}
%Web https://exoplanetarchive.ipac.caltech.edu
The number of confirmed planets from the NASA exoplanets archive\footnote{\href{https://exoplanetarchive.ipac.caltech.edu}{https://exoplanetarchive.ipac.caltech.edu}} is 5347 up to now (05/2023). After cross-matching with K2020, we find 76 confirmed planets in clusters. To study the planet size-age distribution, we select 51 transiting planets with known planet radii. 

The properties of planets and their host stars, e.g. planet radius, orbital periods, effective temperature, surface gravity, and stellar mass are adopted from the table of NASA exoplanets archive. We adopt the age of the host stars from K2020 temporarily, which is completed. The adopted ages from K2020 will be validated via comparison in subsection \ref{subsec:2.2}. 

\subsubsection{K2020 cross-matching with KOIs} 
The number of KOIs from Kepler DR25 \citep{2018ApJS..235...38T} is 8445, including confirmed planets and candidates. After cross-matching with K2020, we find 98 KOIs in clusters. Since some of these KOIs may be False Positives, such as eclipsing binaries in the background of the targets, or physically bound to them which can mimic the photometric signal of a transiting planet. Then, we exclude the sources flagged as False Positive. There are 26 confirmed planets and 17 planet candidates in clusters for Kepler sources. Here, we use the Gaia-Kepler Stellar Properties Catalog \citep{2020AJ....159..280B} to update and obtain the accurate and precise properties of these 43 selected KOIs and their host stars. The age information is also adopted from K2020. 

%and the recent catalog provided by \citealt{2020ApJS..247...28H}？
%sources?. The original catalog of K2 planet candidates from the NASA exoplanet archive does not provide stellar properties such as stellar mass and stellar effective temperature, we use the catalog of \cite{2020ApJS..247...28H} to replenish the missing information. The same as in previous subsections, 
\subsubsection{K2020 cross-matching with K2}
For K2 sources, we use the catalog of K2 including both confirmed planets and candidates from the NASA exoplanet archive to cross-match with K2020. We get only 25 matching sources. After excluding 9 candidates flagged as ``FALSE POSITIVE'', we obtain 9 confirmed planets and 7 planet candidates with planet radius measurements. The properties of planets and planets' host stars are taken from the NASA exoplanet archive. The age information is taken from K2020. 

\subsubsection{K2020 cross-matching with TOIs} %添加关于FA,APC,FP的说明。
Up to 05/2023, there are 6586 TOIs detected by TESS. After cross-matching with K2020, there are 116 TOIs left. There are many False Positives in TOIs \citep{2023AJ....165...17G}. Thus, we select the TOIs carefully. Firstly, we exclude some sources flagged as ``FA'', ``APC'', and ``FP'', which means false alarm, ambiguous planet candidate, and false positive respectively. After exclusion, 68 TOIs are left. Besides, based on publicly available observational notes on ExoFOP\footnote{\href{https://exofop.ipac.caltech.edu/tess/}{https://exofop.ipac.caltech.edu/tess/}}, we also remove some TOIs with comments like ``centroid offset\footnote{The light from nearby eclipsing binaries within 1' may pollute the aperture and cause transit-like signals on the target light curve, especially in a crowded field like open clusters}'', ``V-shaped'', ``Likely eclipsing binary(EB)'', and ``odd-even''. For example, the comment of TOI 1376.01 is ``centroid offset on TIC 190743999 in spoc-s56''. So after removing these TOIs, we finally select 42 TOIs, i.e. 12 confirmed planets and 30 candidates. The properties of these 42 selected TOIs and their host stars are taken from the table of TOIs. The age information is taken from K2020. 

\subsubsection{other sources}
K2020 focuses on the stars with galactic latitude $|b|< 30 ^{\circ}$. Actually, there are many other clusters out of such range, and so do the planets in clusters. We add the planets and planet candidates in the PATHOS project to include more planets at higher galactic latitudes. Table 6 of PATHOS - IV \citep{2021MNRAS.505.3767N}, provides 33 confirmed planets in clusters. Although 11 confirmed planets are repeated selected planets in section \ref{subsec:2.1.1}, 14 of them are with $|b|> 30 ^{\circ}$, and 8 of them are uncross-matched with K2020.
%确定有多少是银纬三十度
Additionally, the PATHOS project has found 90 planet candidates in clusters, which are not included in TOI completely. After cross-matching with K2020, we get the age information of 40 planet candidates with $|b|< 30 ^{\circ}$. Here, we do not include the candidates of the PATHOS project with $|b|> 30 ^{\circ}$, because they do not have the age measurements. \cite{2021MNRAS.505.3767N} provide false positive probabilities for the PATHOS candidates. We remove 8 sources with high false positive probabilities. Most of them are likely eclipsing binaries. Thus, we add 32 PATHOS candidates. The stellar properties are taken from the TESS Input Catalog v8.0 \citep[TIC-8,][]{2019AJ....158..138S}, e.g. stellar mass, stellar effective temperature, and surface gravity.  

\subsubsection{The catalog of transiting planets in star clusters}
After the cross-matching, we check the catalog of transiting planets in clusters and exclude the repeated planets. Finally, there are 73 confirmed planets and 84 planet candidates in 86 clusters. Table \ref{tab:pcl} shows all these planets in 133 planetary systems. Planets detected by different missions are flagged as ``Kepler'', ``K2'', ``TESS'', and ``other''. Here, ``other'' means sources detected by other facilities, e.g. CoRoT \citep{2009A&A...506..287L} and ground-based telescopes. 

This is the largest catalog of the planets and planet candidates in star clusters. Due to space constraints, we only list 10 sources in Table \ref{tab:pcl}. The whole table can be downloaded from the web: \href{https://github.com/astrodyz/UPiC}{https://github.com/astrodyz/UPiC}
%You can find the whole readable catalog in the following link. 

%关于表格的修改。有效数字！

\begin{figure*}
    \centering
    \includegraphics[width=0.95\linewidth]{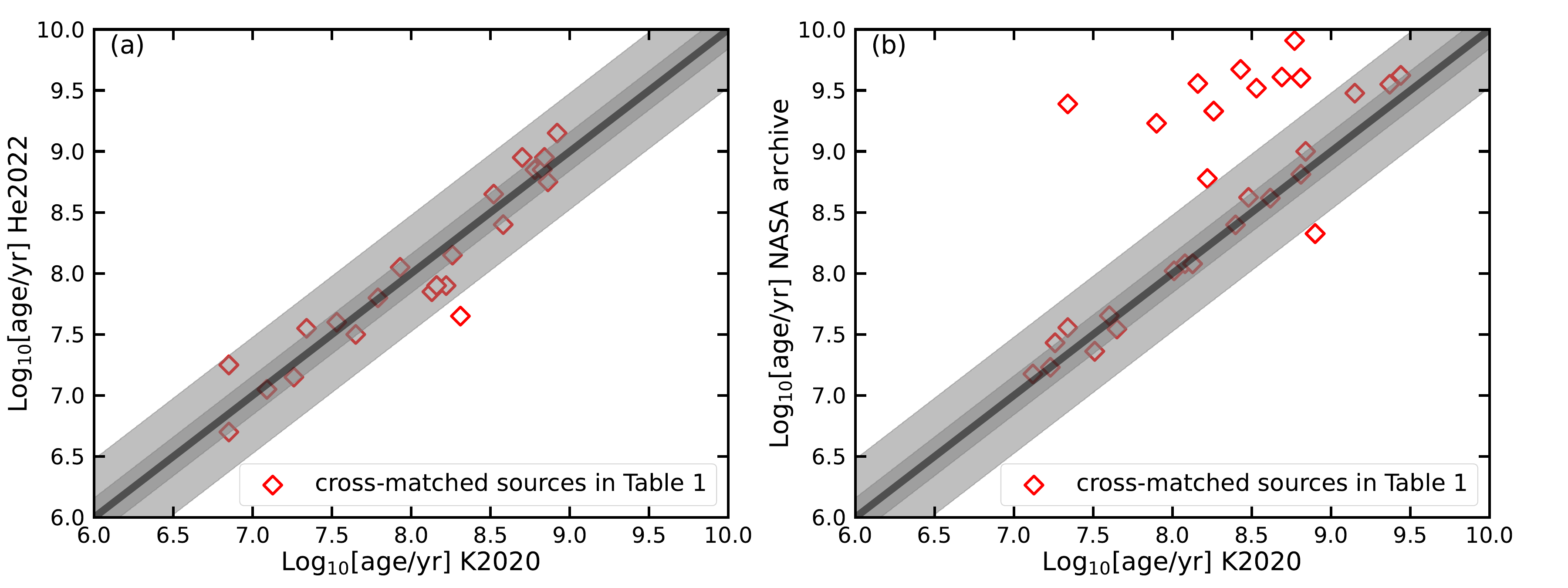}
    \caption{The comparison of age estimation from different catalogs. Panel (a) uses data from the cross-matching catalog of K2020 and He2022. Panel (b) uses data from K2020 and the NASA exoplanet archive. Gray shadows are 1-sigma and 3-sigma areas considering the average uncertainty of age estimation of comoving groups within 1kpc in K2020. A black solid diagonal line is plotted for reference.}
    \label{Figure:age_validation}
\end{figure*}
\subsection{Age validation} \label{subsec:2.2}
%???add a summary about the validation and contamination
%这个部分使用不同星表的数据得到的星团年龄做一下交叉匹配，只要得到的年龄的差距不大，则可以说明我们的结果差距不大，即可。
Since we focus on the age-size distribution of planets in clusters, the accuracy, and precision of age measurements are essential. In this section, we will compare the age measurement of stars from \cite{2022ApJS..262....7H} (hereafter He2020), K2020, and the table of the NASA exoplanet archive to illustrate whether the age estimation in K2020 is robust. 
%把2020年龄的问题讲述清楚。有哪些不确定因素，然后说明虽然整体上K2020的年龄相比于He2022的年龄偏低一些，但是对于两者交叉匹配的样本而言相差很少，95以上都在3sigma范围的误差内。以K2020作为韵致样本的年龄应该是没有问题的。然后说明一些其他的问题，比如说outlier的点看看有没有其他的年龄数据（这个可能需要明天做了）；看一下为啥和NASA archive比对的一些数据差距很大，是不是把一些场星放到里面去了（这样一些年老的场星会被认为是年轻恒星？？！！！如果把这些数据剔除以后样本会如何）。今天晚上写完这一部分+简单的复习。
\cite{2020AJ....160..279K} adopted a neural network called Auriga to robustly estimate the ages of the individual groups they identified. The uncertainty of log(Age) of comoving groups within 1 kpc in K2020 is similar to \cite{2019AJ....158..122K}, i.e. $\sim$ 0.15 dex. In He2022, they use the isochrone fitting to derive the ages of 886 nearby clusters and candidates within 1.2 kpc. 

Although \cite{2020AJ....160..279K} discuss the contamination and demonstrate that the vast majority of the age are well consistent with the results of isochrone fitting, the age estimation in K2020 may still have some systematic biases compared to He2022. On top of that, the difference in identifying clusters may also influence the final age estimation. K2020 used unsupervised machine learning HDBSCAN to identify clusters, while He2022 used DBSCAN. HDBSCAN has a better performance on the data with different density structures than DBSCAN, i.e. prefer to reveal more fine structures. Therefore, both the age estimation and the cluster membership identification will lead to systematic biases.

%Therefore, the catalog present in K2020 is significantly more sensitive to the extended structure. ???However, background co-moving stars will bring some contamination, i.e. 5\% - 10\% of the population inside the identified groups consisting of field stars with kinematics similar to those of the underlying cluster(\cite{2019AJ....158..122K}). 
%修改图片对比度要增加。
%Although the age estimation of these 81,479 stars in K2020 tends to be older than that in He2022 and the diversity is large, these two catalogs are nearly consistent with each other for the age of those 26 planets' host stars in clusters. , and Gaia 3218880217994699136 and 3218615166972886016 in NGC 2112,The blue line shows where the age estimation in K2020 equals He2022. Blue and purple shadows are 1-sigma and 3-sigma areas considering the average uncertainty of age estimation in K2020. While for NGC 2112, several works (e.g. \cite{2008MNRAS.386.1625C,2013A&A...558A..53K,2017Ap.....60..173H,2021MNRAS.504..356D})present an age larger than 1 Gyr, which is closer to the estimation in He2022. Thus, we adopt the age estimation of NGC 2112 in He2022, 1.4 Gyr. 
Then, we cross-matched the catalog of K2020 and He2022 to present whether the bias of age estimation in K2020 is non-negligible compared to other age sources. In panel (a) of Figure \ref{Figure:age_validation}, red hollow dots are the 36 planets/candidates in 22 clusters that both have the age estimation in K2020 and He2022. Only one planet candidate has a large difference of age estimation, out of the 3 sigma limit, i.e. the PATHOS 64 in King-6. Therefore, we assume that the majority of the age estimation in K2020 is relatively robust. To validate the ages of this cluster, we refer to the ages from other previous works. In K2020, the estimation of the age of King-6 is 204$^{+91}_{-63}$ Myr which is consistent with the previous result of \cite{2002AJ....123..905A}(250$\pm$50 Myr), while in He2022, the age of King-6 is 44 Myr without uncertainty. Thus, we adopt 204$^{+91}_{-63}$ Myr as the final age of King-6. 

%i.e. the age estimation in K2020 tends to be larger
%这里还需要重新修改一下。这个原比较多使用表格还是？先整理再说。
Additionally, we also compare the age of the K2020 and the NASA exoplanet archives. In panel (b) of Figure \ref{Figure:age_validation}, there are 29 planet host stars(44 planets) in our catalog with both ages from K2020 and NASA exoplanet archives. Eleven host stars are out of the 3 sigma limit in the age measurement, i.e. CoRoT-22, HATS-47, HD 110113, KELT-20, Kepler-1062, Kepler-1118, Kepler-1502, Kepler-411, Kepler-968, TOI-1937 A, and TOI-4145 A. All the ages of these planets' host stars from NASA exoplanet archives are much larger than the ages from K2020. E.g. the age of CoRoT-22 is 3.3$\pm$2.0 Gyr in \cite{2014MNRAS.444.2783M}, while in K2020 is 339$\pm$ 100 Myr; the age of HATS-47 in \cite{2020ApJS..247...28H} is 8.10$^{+2.90}_{-4.30}$ Gyr, while in K2020 is 589$^{+243}_{-171}$ Myr; the age HD 110113 \cite{2021MNRAS.502.4842O} is 4.0$\pm$ 0.5 Gyr, while in K2020 is 645$^{+245}_{-178}$ Myr.

Because the individual age estimation of these stars depends on the models and methods, which are inhomogeneous in the NASA exoplanet archive. Strictly, we use different ways to validate the ages of the eleven host stars.

%we do not use these data in the following analysis. For example,  The age difference between these three stars are nearly one order of magnitude.  
%, Kepler-1062 is $\sim$ 21 Myr in Stephenson 1 (53.8 Myr, \cite{2009MNRAS.399.2146W}) and 
Firstly, some stars may have several($\geq$ 3) age measurements, which we can evaluate through majority voting. For instance, the age estimation of KELT-20 is $\le $0.6 Gyr according to \cite{2017AJ....154..194L}. However, according to \cite{2018A&A...612A..57T}, KELT-20 is   
200 $^{+100}_{-50}$ Myr, which is consistent with K2020's result, i.e. 166$^{+58}_{-43}$ Myr. The age estimation of Kepler-1118 from \cite{2016ApJ...822...86M}(4.07 Gyr) is nearly ten times of that in K2020(490$^{+155}_{-118}$ Myr). We strengthen that \citeauthor{2016ApJ...822...86M} set the age prior to 1-15 Gyr, which means the stellar age in their catalog is artificially larger than 1 Gyr. As the host cluster of Kepler-1118, NGC 6866 has the age of 705$\pm$ 140 Myr from \cite{2014AJ....147..139J}). Therefore, we adopt the age in K2020 for KELT-20 and Kepler-1118. 

%from K2020 is 490$^{+155}_{118}$ Myr, which is consistent with evaluate through majority voting
Secondly, if stars do not have independent and consistent age measurements, we estimate the age via the gyrochronological relation. Kepler-411 is a special case that three age measurements are significantly different. \cite{2019A&A...624A..15S} use gyrochronological relation (\cite{2007ApJ...669.1167B}) and estimate an age of 212 $\pm$ 31 Myr, K2020 estimate an age of 794$^{+302}_{-219}$ Myr, and \cite{2016ApJ...822...86M} estimate an age of 2.69$^{+2.67}_{-1.10}$ Gyr.) To validate the age of Kepler-411, we adopt the new gyrochronological relation, which includes the empirical mass dependence of the rotational coupling timescale, developed by \cite{2020A&A...636A..76S}. The stellar rotation period of Kepler 411, 10.4 days, is taken from \cite{2013ApJ...775L..11M}. Then, we estimate an age of 770 Myr for Kepler 411 system which is consistent with the age estimation in K2020. Thus, we adopt 794$^{+302}_{-219}$ Myr as the age of Kepler-411. 

The same with Kepler-411, we use the rotating period and stellar mass, and obtain the ages of HATS-47, HD 110113, Kepler-1062, and Kepler 968, through the new gyrochronological relation, i.e. $< 0.1$ Gyr, $\sim$ 3 Gyr, $\sim$ 1.3 Gyr, and $\sim$ 0.7 Gyr, respectively. These age estimations are significantly different from K2020 (i.e. 589$^{+243}_{-171}$ Myr, 645$^{+245}_{-178}$ Myr, 21$^{+4}_{-4}$ Myr, and 181$^{+52}_{-41}$Myr, respectively). We speculate HATS-47, HD 110113, Kepler-1062, and Kepler 968 may be the contaminating stars in star cluster identification. They may be field stars having similar kinematic properties compare to the comoving stellar groups in their proximity coincidently. Therefore,  we exclude these three potential contaminating sources in cluster identification. We also remove HATS-47 because it does not have a convincing age measurement. 

%do not have more than three age measurements and without additional rotation period measurements, we exclude them. For example, 
Thirdly, for CoRoT-22, Kepler-1502, TOI-1937 A b, and TOI-4145 A b which have inconsistent age measurements and lack of stellar rotation measurements, we can hardly validate their age. Additionally, \cite{2023ApJS..265....1Y} suspects that TOI-1937 A and TOI-4145 A may be the field star because of the poorly constrained cluster membership identification. Therefore, we directly exclude these four systems. 

%关于锂元素丰度的讨论?一笔带过还是？似乎使用Gaia-ESO survey也不适合，因为没有找到很多相应的恒星。不知道这样论述可不可以。后面加上Gaia-ESO的acknowledge和引用文章

%\textbf{In Addition to gyrochronological relation, lithium abundance is another method of age measurement, especially for stars younger than 1 Gyr. Here, we use the software --``Empirical AGes from Lithium Equivalent width'' \citep[EAGLES,][]{2023MNRAS.523..802J}\footnote{\href{https://github.com/robdjeff/eagles} {https://github.com/robdjeff/eagles}} to check some age measurements. We cross-match our planet hosts with the Gaia-ESO survey. However, no matched planet host stars are found. We only check the ages of three open clusters both in our planet host samples and the Gaia-ESO survey, i.e. NGC 2516, IC 2602, and Vela OB2, whose lithium ages are consistent with cluster ages. More stars with lithium abundance based on spectroscopic surveys will help our age validation.}

%check the number 

To sum up, we validate the age measurement of 70 planets/candidates in star clusters, obtain more convinced ages of three host stars via either literature or the new gyrochronological relation, and exclude eight planetary systems without convincing age estimations. If we assume those eight host stars are field stars, the contamination rate of our catalog is about 6\%, which is consistent with that in \cite{2019AJ....158..122K}, i.e., 5\%-10\%.  

%exclude 3 planets i.e. CoRoT-22 b, HATS-47 b, and HD 110113 b with different age measurements. 
%??? table + validated age ?
\subsection{Sample cut} \label{subsec:2.3}
In section \ref{subsec:2.2}, we obtain 63 planets and 84 planet candidates in star clusters with relatively robust age estimation. We aim to obtain the planet radius evolution, i.e. planet radius--age distribution. The accuracy of planet radius and age measurement will significantly influence our results. Therefore, we need to do the sample cut to minimize the influence of observational biases. 

%In section 2.2, we do the age validation and show that most of the age estimation in K2020 is consistent with other measurements except for three confirmed planets. All these planets and planet candidates are detected through the transit method. 
Here, we list the steps of sample cut in Table \ref{tab:cut}. Without the mass measurement, we can hardly determine whether the planet candidates are planets or brown dwarfs. Planet candidates with large radii are unlikely planets. Thus, we exclude 23 planets/candidates, with $R_{\rm p} > 2.5 \,\rm R_{\rm J}$ (the same criteria described in \cite{2021MNRAS.505.3767N}). Brown dwarfs with larger masses can induce the motion of the photon center. \cite{2020AJ....159..280B} suggest that stars with high re-normalized unit-weight error (RUWE$>$1.4), are likely to be binaries. \textbf{Although, a few confirmed astrometric planets from Gaia \citep{2023A&A...674A..10H} have RUWE$>$1.4), most stars with confirmed planets are below this threshold.} Thus, we adopt the criteria, RUWE$<$1.4, to exclude 9 planet candidates in potential binary systems. 
%\textbf{(add ruwe discussion in here or later discussion)}
Since planets with poor radius measurements may contaminate the results, we exclude 7 samples with relative radius errors larger than 50\%. Due to the precision of TESS and the stellar noise of young stars, small planets detected by TESS and planets around young stars are less complete. Thus, we need to constrain the lower limits of the planet radius to exclude the bias of completeness. As shown in Appendix \ref{appendix:a}, planets with radius $R_{\rm p}>2\rm R_{\oplus}$ and period $P<20$ days \textbf{could be detected} (SNR$>$7.1) via both Kepler and TESS. Thus, we cut the sample via $R_{\rm p}>2 \rm R_{\oplus}$ and $P<20$ days. 

After the sample cut, there are 66 planets/candidates left. In the next section, we mainly use the sample to do the analysis.

%这些数字可能都需要修改。
%Young stars with higher stellar noise may also contaminate the transit signal. Small planets around young stars are difficult for TESS to detect. On top of that, TESS observes each star in a shorter time-spanning than Kepler. Different facilities have different detection efficiency. We should do the data reduction on planet radius and orbital periods to minimize this observation bias. 

%Last, we adopt the criteria that orbital periods are shorter than 20 days and a planet radius is larger than 2 Earth radii (See more details in Appendix \ref{appendix:a}), and select 62 planets and planet candidates. In the following, we mainly use these 62 samples to do the analysis.

\begin{deluxetable}{ccc}
\centering
\tabletypesize{\scriptsize}
\tablewidth{0pt} 
\tablenum{2}
\tablecaption{Sample Cut of Planets/Candidates in clusters \label{tab:cut}}
\tablehead{
\colhead{Criterion}     &\colhead{Planets}      & \colhead{Planet candidates}\\
}
\startdata 
The whole number & 73 & 84 \\
Age validation & 63 & 84 \\
$R_{\rm p} < 2.5\, \rm R_{\rm J}$ & 62 & 62 \\
RUWE $< 1.4$ & 62 & 53 \\
$\sigma_{\rm R_{\rm p}}/R_{\rm p}< 0.5$ & 58 & 50 \\
$P < 20$ days & 37 & 44 \\
$R_{\rm p}>2 \rm R_{\oplus}$ & 30 & 36 \\
\enddata
%\tablecomments{Sample Cut of Planets in clusters}
\end{deluxetable}

\section{Planet Radius--Age Distribution} \label{sec:3}
\begin{figure*}
    \centering
    \includegraphics[width=1\linewidth]{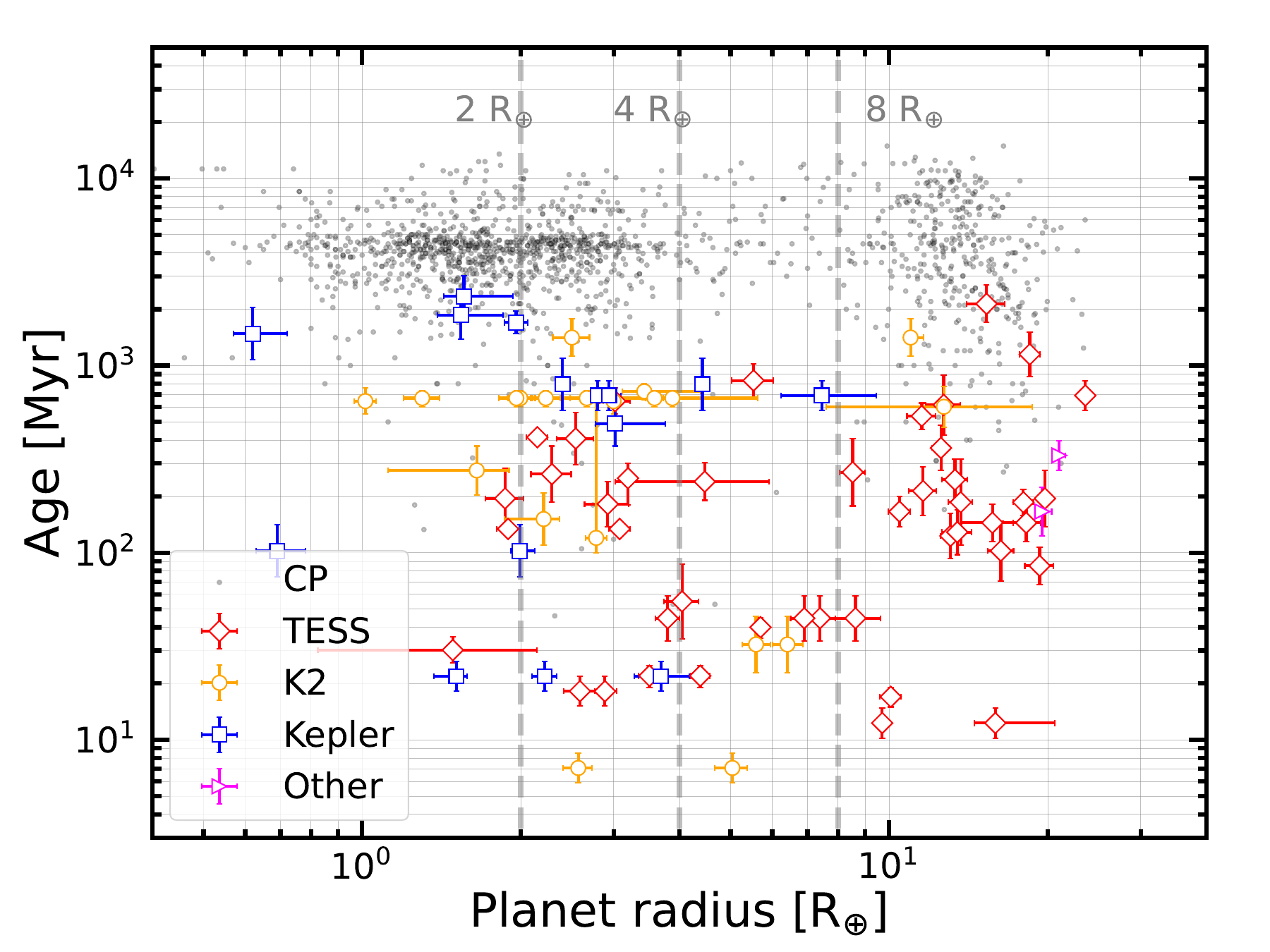}
    \caption{The planetary radius -- age distribution of 37 planets and 44 planet candidates in star clusters(Table \ref{tab:cut}). Different colors show the planets in clusters detected by different missions. Red, yellow, blue, and purple symbols are planets/candidates detected by different missions, i.e. Kepler, K2, TESS, and other ground-based telescopes, respectively. The gray dots are confirmed transiting planets whose host stars have age measurements. }
    \label{Figure:sizeage}
\end{figure*}

\subsection{Planet Radius--Age Diagram} \label{subsec:3.1}
Figure \ref{Figure:sizeage} shows the planetary size -- age distribution of 37 planets and 44 planet candidates in star clusters(15 planets/planet candidates with $R_{\rm p}<2 \, \rm R_{\oplus}$ and 66 planets/planet candidates with $2 \, \rm R_{\oplus}< R_{\rm p}<2.5 \, R_{\rm J}$). 

%Different colors show the planets in clusters detected by different telescopes. Red, yellow, blue, and purple symbols are planets or planet candidates detected by TESS, K2, Kepler, and other ground-based telescopes respectively. ??Panel (a) and (b) are different in age estimation. In panel (a), we choose the age estimation of K2020. The black dots are other confirmed planets around field stars that have age estimations. 

%?item SubN means a sub-Neptune-sized planet
%[2]SubS means a sub-Saturn-sized planet(); 
%[3]HJ is Jupiter-sized planet(8 R$_{\oplus}$ < R$_{p}$ < 2.5 R$_{J}$). 
Here, we classify planets into three groups by size for the sake of simplicity:

\begin{itemize}
    \item Sub-Neptunes, i.e. planets of $2 \,\rm R_{\oplus}<  R_{\rm p} < 4\, \rm R_{\oplus}$,
    \item Sub-Jupiters, i.e. planets of $4\,\rm R_{\oplus} < R_{
    \rm p} < 8 \,\rm R_{\oplus}$,
    \item Jovian planets, i.e. planets of $8\,\rm R_{\oplus} < R_{\rm p} < 2.5\, \rm R_{\rm J}$.
\end{itemize}

There are only 5 Jovian planets younger than 100 Myr, while dozens beyond 100 Myr. So, it seems that there is a gap in the planet radius--age diagram for the Jovian planets younger than 100 Myr.
Additionally, there may be another gap for Sub-Jupiters with ages between 50--200 Myr. Before 50 Myrs, there are several Sub-Jupiters, while between 50--200 Myr, the number of Sub-Jupiters declines to nearly none. On top of that, there are nearly no Sub-Neptunes between 50--100 Myr.

%这一段怎么总结一下逻辑，引出我们需要对相对比例的研究。
%有没有可能是100Myr的时候的星团本身就比较少

Therefore, it seems that all of the planets disappear between 50--100 Myr. However, due to the small number of planets/candidates(i.e. the large statistical error), whether the gap is real can not be easily demonstrated. In order to avoid observational bias, in the next subsection, we will take into the age error and the radius error of the planets to obtain the time-dependent relation for the proportion (instead of the number) of different-sized planets in star clusters.

\begin{figure}
    \centering    
    \includegraphics[width=1\linewidth]{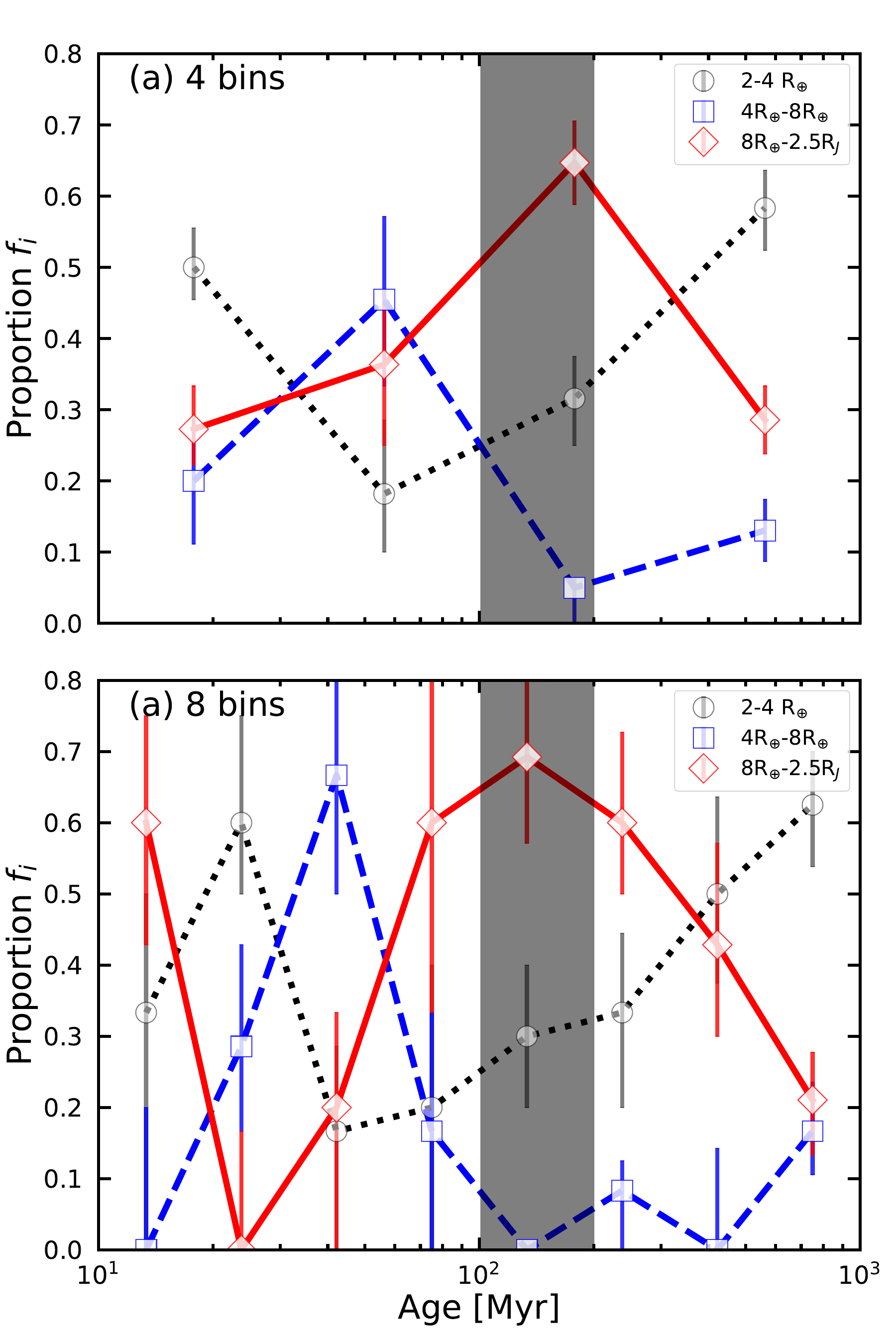}
    \caption{The time-dependent relation of the proportions of planets(in star clusters) of different sizes. Different colors show planets of different sizes. Panel (a) and (b) are different in numbers of age bins, i.e. 4 age bins between 10 to 1000 Myr under log scale in panel (a), and 8 age bins in panel (b).}
    \label{Figure:cluster}
\end{figure}

\begin{figure*}
    \centering    
    \includegraphics[width=1\linewidth]{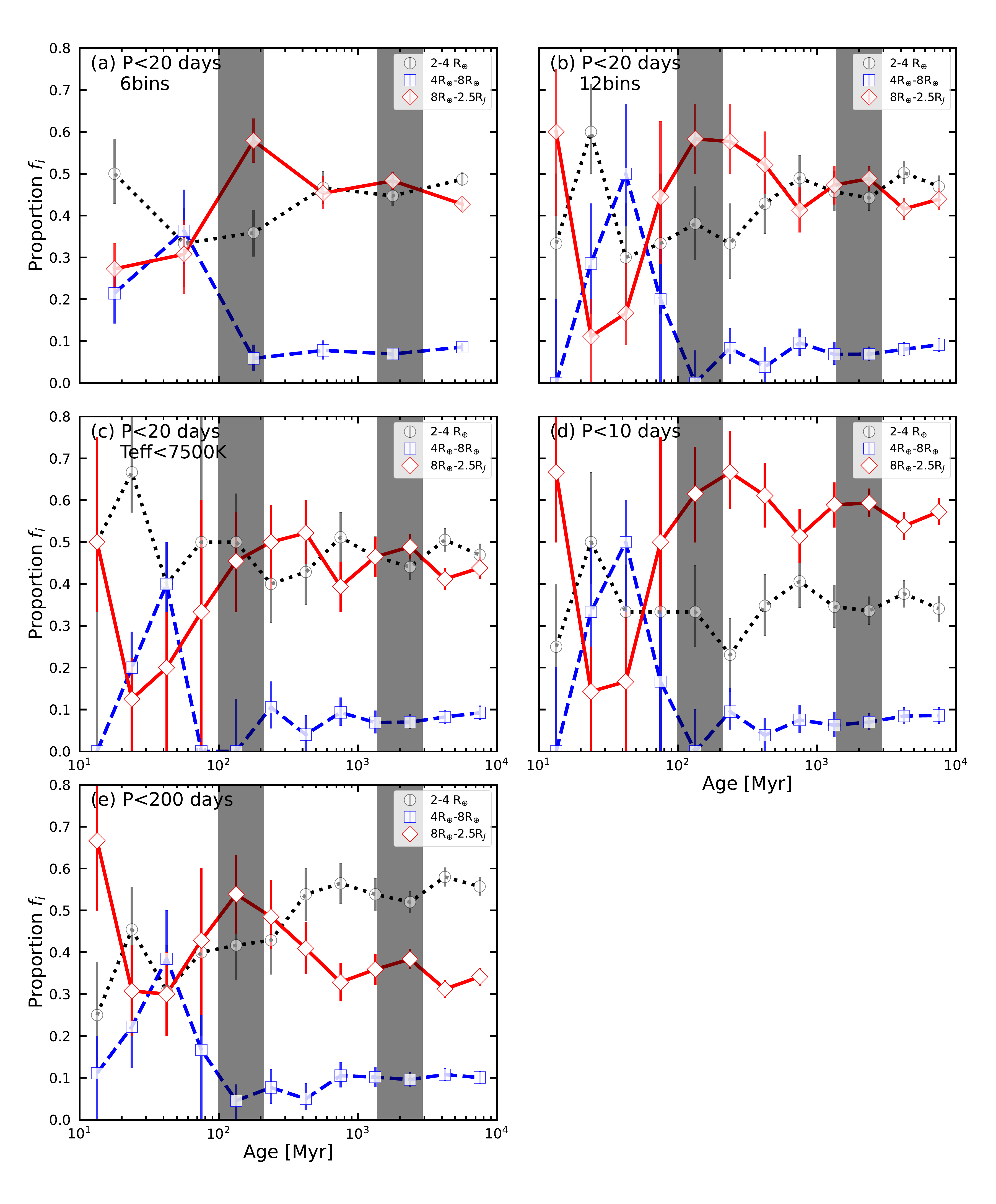}
    \caption{The time-dependent relation of the proportions of planets(both in star clusters and probably around field stars) of different sizes. Different colors show planets of different sizes. Panel (a) and (b) are different in the number of age bins, i.e. 6 age bins between 10 to 10000 Myr under log scale in panel (a), and 12 age bins in panel (b). We add a cut of effective temperature in panel (c) and an additional period cut in panel (d), i.e. $P<10$ days. For panel (e), we show the result for planets with $P<200$ days as a comparison. Two shadow regions around 100 Myr and 2 Gyr are over-plotted to emphasize two typical timescales. }
    \label{Figure:relative}
\end{figure*}

\subsection{The Evolution of Planet Radius}\label{sec:3.2}
\textbf{In this section, we will show our main results of the time-dependent relation of different-sized planets in section \ref{sec:3.2.1} and some influence of contamination in section \ref{sec:3.2.2}}
%\subsubsection{The proportions of planets with different sizes and ages}
\subsubsection{Main results}\label{sec:3.2.1}
To investigate the time-dependent relation of different-sized planets in star clusters. We defined the proportions of planets with different sizes and ages. To determine the proportions and the uncertainties, we randomize the ages and radii of all the planets 100,000 times, assuming the Gaussian distribution. Then, we can obtain the proportions of each time, denoted as $f_{\rm i}$, via the formula: 

\begin{equation}
    f_{\rm i} = \frac{N_{\rm i}}{N_{\rm SubN}+N_{\rm SubJ}+N_{\rm J}},
\end{equation}
where $N_{i}$ is the number of planets in star clusters with different sizes, i.e. $N_{\rm SubN}, N_{\rm SubJ}$ or $N_{\rm J}$ corresponding to Sub-Neptunes, Sub-Jupiters, and Jovian planets in section \ref{subsec:3.1}, respectively. After calculating 100,000 times, we obtain the distribution of $f_{\rm i}$ and adopt the lower limits, median values, and the upper limit, according to the 16, 50, and 84 percentiles of $f_{\rm i}$, respectively. 
%所有的符号fJ改成统一格式
The results are as shown in Figure \ref{Figure:cluster}. Note the age is cut by $\leq$ 1 Gyr because most of the selected planets/candidates in star clusters are younger than 1 Gyr(Figure \ref{Figure:sizeage}). Panel (a) and (b) are different in numbers of age bins, i.e. 4 and 8 age bins between 10 to 1000 Myr under the log scale, respectively. Both two panels show that the proportion of Jovian planets (red diamonds, $f_{\rm J}$) increases before 100 Myr, and then declines after 200 Myrs, i.e. a peak occurs between 100 and 200 Myr. The proportion of Sub-Jupiters (blue squares, $f_{\rm SubJ}$) declines around 100 Myr. The proportion of Sub-Neptunes (gray circles, f$_{\rm SubN}$) shows a clear increase after 100 Myr. Here, we also consider the Poisson error because of the small number of planets in each bin, which has similar features to Figure \ref{Figure:cluster}(See Figure \ref{Figure:poisson} in Appendix \ref{appendix:b}). 

To extend the evolution of planets/candidates older than 1 Gyr, we calculate the proportion of planets both in star clusters and around field stars, with different sizes. Here, we add $\sim$ 871 confirmed planets with age measurements from the NASA Exoplanet Archive. These confirmed planets share the same cut with planets/candidates in star clusters, i.e. 2$\rm R_{\oplus}<R_{\rm p}<2.5 \rm R_{\rm J}$, and $P<20$ days. Using the same estimating procedures as those in Figure 
\ref{Figure:cluster}, we obtained the proportion varying with age, as shown in Figure \ref{Figure:relative}. 

%Each random, we can get one list of proportions of different-sized planets. After 100,000 times, we can get the distribution of the proportions of planets with different sizes and ages. The errors of the proportion are according to the 16\%, 50\%, and 84\% percentile values of every array of the proportion. Then, we obtain the proportion of planets of different sizes changing with age.  
%move

%?As shown in Figure \ref{Figure:relative}, there are two shadowed regions, . One is around 100 Myr, the other is around 2 Gyr. Note, the planets in star clusters are dominated within 300 Myr, while the planets around field stars become dominated beyond 1 Gyr. 
%, in panel (b) of Figure \ref{Figure:sizeage}, after a rapidly increasing around 100 Myr, $f_{\rm J}$ then declines a bit after 200 Myr, i.e., a peak between 100 and 200 Myr. Besides, $f_{\rm SubJ}$ rapidly decreases around 100 Myr and then remains at low value $\sim$ 0.1. Additionally, 
In the panel (a) and (b) of Figure \ref{Figure:relative}, the proportion of Jovian planets (red diamonds, $f_{\rm J}$), Sub-Jupiters (blue squares, $f_{\rm SubJ}$), and Sub-Neptunes (gray circles, $f_{\rm SubN}$) show the similar time-dependent relation within 1 Gyr to that in Figure \ref{Figure:cluster}. I.e. $f_{\rm J}$ reaches maximum between 100 and 200 Myr, $f_{\rm SubJ}$ rapidly declines around 100 Myr, and $f_{\rm SubN}$ increases after 100 Myr. Because Figure \ref{Figure:relative} has a longer time span than Figure \ref{Figure:cluster}, there are more substructures in Figure \ref{Figure:relative}. For example, all of the panels in Figure \ref{Figure:relative} show a tiny bump of $f_{\rm J}$ around 2 Gyr, which is anti-correlated with $f_{\rm SubN}$, i.e. a small dip of $f_{\rm SubN}$ around 2 Gyr. These two timescales, i.e. 100 Myr and 2 Gyr (gray shadow regions), may correspond to different planet formation environments (see discussion in \ref{subsec:4.2}). Because the majority of planets younger than 300 Myr are in star clusters, while most of the planets older than 1 Gyr are around field stars.

To illustrate that our results are robust, we should exclude the influence of some other stellar parameters. For example, some planet host stars are very hot, i.e. $T_{\rm eff} > 7500$ K, especially for the candidates detected by TESS. For main-sequence stars, hotter stars usually have larger stellar radii than cooler ones. Therefore, the transit method tends to find larger planet candidates around hotter stars. In panel (c), we add another criterion for planets/candidates in star clusters and around field stars, i.e. $T_{\rm eff}< 7500$ K. Although the proportion of Jovian planets $f_{\rm J}$ around 200 Myr is smaller than that in panel (b) because of the additional sample cut, $f_{\rm J}$ still continuously increases between 100 and 400 Myr, i.e., the peak moves backward to around 400 Myr. The proportion of $f_{\rm SubJ}$ rapidly decreases around 100 Myr, then remains at $\sim$ 0.1 after 100 Myr, similar to that in panel (b). $f_{\rm SubN}$ does not show an obvious increase/decrease after 100 Myr. 

The widely used definition of Hot Jupiters(HJs) is Jupiter-sized planets within 10 days \cite{2018ARA&A..56..175D}. Here, in panel (d), as a comparison, we also show the results of the conventional hot planets within 10 days. For Jovian planets and sub-Jupiters, panel (d) shows similar results to panel (b). Therefore, in the following, we call the Jovian planets within 20 days as HJs for simplicity (if without additional annotation). For sub-Neptunes, the increasing tendency after 100 Myr is ambiguous. 

In panel (e), we show the results of the time-dependent relation of planet radius for planets within 200 days. Because including some warm planets, the increment of $f_{\rm J}$ around 100 Myr in panel (e) becomes less than that in panel (b). As the majority of the warm planets are sub-Neptunes, $f_{\rm SubN}$ in panel (e) is systematically higher than that in panel (b) after 100 Myr. In turn, the $f_{\rm J}$ in panel (e) is systematically lower than that in panel (b) after 100 Myr. The time-dependent relation of $f_{\rm SubJ}$ in panel (e) is similar to other panels. 

In Figure \ref{Figure:relative}, we can not find an obvious increase trend of the proportion of sub-Neptunes within 1 Gyr as shown in Figure \ref{Figure:cluster}. Because it seems sensitive to the parameter cut.

%% addition influence
\subsubsection{Influence of some contamination} \label{sec:3.2.2}
\textbf{To test the robustness of our statistical results, we also check some other influences, e.g. the criteria of planet radius cut and the potential false positives in planet candidates.} 

\textbf{Firstly, we will discuss the criteria of planet radius cut. Young giant planets with expanded radius may still go through the contraction, which means the exclusion of young candidates with $R_{p} > 2.5 R_{J}$ may underestimate the $f_{J}$. We checked the 23 candidates with $R_{p} > 2.5 R_{J}$ in section \ref{subsec:2.3} and found that most of them are around 100 Myr. If we consider these candidates, it will enhance the peak of $f_{J}$ between 100 and 200 Myr.}

\textbf{Secondly, to avoid the influence of potential false positives, we only use confirmed planets to drive the time-dependent relation of planet radius. The results are shown in Figure \ref{Figure:confirmed} in Appendix \ref{appendix:c}, which is similar to that of containing planet candidates. Therefore, our statistical results are robust.}

To summarize section \ref{sec:3.2}, we obtain the time-dependent relation of planet radius for planets/candidates in star clusters and around both cluster members and field stars, i.e.  
\begin{itemize}
\item The proportion of Jovian planets $f_{\rm J}$ increases around 100 Myr and reaches a maximum between 100 Myr and 200 Myr, which is mainly attributed to the HJs in star clusters. The tiny bump of $f_{\rm J}$ around 2 Gyr is attributed to the HJs around field stars. 
\item The proportion of Sub-Jupiters $f_{\rm SubJ}$ declines rapidly around 100 Myr, then remains at a low value. The declination of $f_{\rm SubJ}$ is mainly attributed to the hot Sub-Jupiters in star clusters.
\end{itemize}

%, i.e. planetary atmospheric mass loss driven by strong photo-evaporation and flyby induced high-e migration in star clusters. 

\section{Constraints of Hot giant planets formation timescale} \label{sec:4}
Based on the statistical results above, we try to explain or constrain the timescales of hot giant planets' formation mechanisms in star clusters. Here, the hot giant planets mean HJs and hot Sub-Jupiters (or hot Neptunes). 
%\subsection{The formation timescale of HJ}

There are several formation scenarios of HJ, i.e. in-situ formation when disk mass is large, or ex-situ formation then undergoing disk migration or high-e migration. The time scales of the first two HJ formation scenarios are mainly limited by the lifetime of the gas disk, which is typically $\sim$10 Myr. Therefore, if the in-situ formation and disk migration are the dominant channels of HJ formation, the number of HJs will not significantly change after 10 Myrs. However, Figure \ref{Figure:cluster} and \ref{Figure:relative} show that the proportion of HJs($f_{\rm J}$) has an obvious increment around $\sim$ 100  Myrs, which is probably attributed to the high-e migration.  

Note, we do not exclude the possibility of HJs forming through the in-situ formation and disk migration. However, with a lack of clusters younger than 10 Myr, we can hardly constrain the in-situ formation mechanism and the fraction of such planets. 

In the following discussion, we mainly focus on the increment of $f_{\rm J}$ and the rapid declination of $f_{\rm SubJ}$ around 100 Myrs in star clusters. More specifically, in section \ref{subsec:4.1}, we estimate the timescales of flyby-induced high-e migrations in star clusters, using typical parameters. In section \ref{subsec:4.2}, we try to explain the tiny bump of HJs, as well as the small dip of sub-Neptune. The Hot-Neptune desert is also discussed in section \ref{subsec:4.3}. \textbf{Some preliminary results of warm Jupiters are shown in section \ref{subsec:4.4} to support flyby-induced high-e migrations in star clusters.}

%Although the high-e migration timescale is model dependent with a large parameter space\cite{2018ARA&A..56..175D} for field stars, 
%The formation timescale of HJ induced by  is model dependent, whose parameter space is very large \cite{2018ARA&A..56..175D}. In the following, we will further discuss the high-e migration in star clusters to constrain the formation timescale of HJs. 
%Figure \ref{Figure:sizeage} shows that there are several hot planet candidates with Jupiter-size around 10 Myrs. On top of that, in Figure \ref{Figure:relative}, the proportion of HJs seems to decline after 10 Myrs. But we can not determine whether the declination is true. Because the declination of the proportion of HJs may just be the observational bias due to the data limitation. In the next discussion, we will focus on the increase of HJs proportion around 100 Myrs. 

\subsection{Flybys induced high-e migration in open clusters within 200 Myr}\label{subsec:4.1}
%In this section, we will review the previous models on HJs formed through high-e migration. We want to discuss how HJs can form around 100 Myr through stellar flyby-induced high-e migration, and give some constraints on the parameters of star clusters if this formation scenario of HJs is dominant. 

%这一段可能说的太啰嗦了。
Recently, several observation works have shown environments in clusters can influence planet formation and evolution (\citealt{2020Natur.586..528W,2021AJ....162...46D}). In star clusters, especially dense clusters, close stellar flybys may occur frequently. A series of theoretical works have shown that the HJ formation can be triggered by stellar flybys in star clusters (\citealt{2020ApJ...905..136W,2022MNRAS.509.5253W,2020MNRAS.499.1212L,2021ApJ...913..104R,2023MNRAS.518.4265L}). Similar to previous works, we consider hierarchical planet systems with both Jovian planets and an outer companion (e.g. a cold giant planet, sub-stellar, or stellar companion). The high-e migrations of the Jovian planet induced by flybys can be described as follows. During a close flyby event, a flyby star exchanges the angular momentum with the outer companion and excites its eccentricity and inclinations. Consequently, the eccentricities of the Jovian planets will be highly excited through the von Zeipel–Lidov–Kozai mechanism \citep[ZKL,][]{1910AN....183..345V,1962AJ.....67..591K,1962P&SS....9..719L}. Finally, tidal circularization leads to the inward migration of the Jovian planet.

There are three factors determining the formation timescale of HJs under flybys-induced high-e migration in star clusters, i.e. the timescale of the close flyby ($\tau_{\rm flyby}$), the ZLK mechanism ($\tau_{\rm ZKL}$), and the tidal circularization ($\tau_{\rm tidal}$). \cite{2021MNRAS.508.3710R} demonstrates that an effective stellar flyby means flyby with small periastron $q$, which can trigger the ZKL oscillation successfully, and subsequently excite the high-eccentricity of the  innerJovian planet. Here, we combine the equation (17),(18), and (19) in \cite{2021MNRAS.508.3710R}, to get an estimation of $\tau_{\rm flyby}$, the timescale of an effective flyby, i.e., 

\begin{equation}
    \tau_{\rm flyby} = \left(\frac{10^{3} \, \rm pc^{-3}}{n_{*}}\right) \left(\frac{2 \, \rm M_{\odot}}{M_{\rm tot}}\right) \left(\frac{50 \, \rm AU}{a_{\rm out}}\right)\left(\frac{\sigma_{*}}{1 \, \rm km/s}\right) \rm Gyr,
\end{equation}
where $n_{*}$ is stellar density in clusters, $M_{\rm tot}$ is the total mass of the hierarchical three-body system, $a_{\rm out}$ is the semi-major axis of the outer companion, and $\sigma_{*}$ is the velocity dispersion of star clusters. Here, we assume a solar-Jupiter system plus a solar-like companion, i.e. $M_{\rm tot} \sim 2 \rm \, M_{\odot}$. Other parameter setting are as follows, $n_{*}=10^{4}$ stars pc$^{-3}$, $a_{\rm out}$ = 50 AU, and $\sigma_{*}$ = 1 km/s. The setting of $\sigma_{*}$=1 km/s and $a_{\rm out}$ = 50 AU are the same as \cite{2021ApJ...913..104R}. 

\textbf{In the following, we describe the reason for the parameter setting of stellar density. According to the hierarchical star formation scenario \cite{2012MNRAS.426.3008K} and recent Gaia DR2 observation \cite{2021A&A...645L...2A}, only a small amount of stars ($<30\%$) have originated from in the bound clusters. I.e. most stars have originated from relatively low-mass stellar groups. Unlike bound open clusters, which probably span over several hundreds of million years, low-mass stellar groups in the solar neighborhood with filamentary substructures are usually younger than 100 Myr \cite{2022ApJ...931..156P}. They may disperse and become associations, e.g. some of the associations might be originally linked to open clusters \cite{2021ApJ...915L..29G}. Some earlier works (e.g. \citealt{2010ARA&A..48...47A,2011MNRAS.411..859M,2013ApJ...769..150C}) assume that high stellar density can only exist in the high-mass clusters. However, the recent work \cite{2021ApJ...921...90P} revealed that low-mass clusters share a similar flyby frequency with high-mass clusters at least in the early stage of cluster evolution. Thus, it's reasonable to assume the typical density of star clusters in their early stage as 10$^{4}$ stars pc$^{-3}$. Then, the timescale of an effective flyby $\tau_{\rm flyby}$ is about 100 Myr.}

The ZLK timescale($\tau_{\rm ZKL}$) can be estimated as \cite{2022MNRAS.509.5253W}: 
\begin{equation}
    \tau_{\rm ZKL} = P_{\rm in}\left(\frac{M_{\rm tot,in}}{M_{\rm out}}\right)\left(\frac{a_{\rm out}}{a_{\rm in}}\right)^{3}(1-e^{2}_{\rm out})^{3/2}
\end{equation}
where $P_{\rm in}$ is the orbital period of the inner Jovian planet, $M_{\rm tot,in}$ is the total mass of the Sun-Jupiter system, $M_{\rm out}$ is the mass of outer companion, and the $e_{\rm out}$ is the eccentricity of the outer companion. If we assume that the inner Jupiter form outside the water ice line around 2.7 AU, i.e. $a_{\rm in}\sim$ 2.7 AU and $P_{\rm in}\sim$ 4.4 yr, the typical ZLK timescale $\tau_{\rm ZKL}$ is $\sim$ 0.3 Myr, which is much shorter than $\tau_{\rm flyby}$.

According to Figure 2 in \cite{2021ApJ...913..104R}, an effective flyby can successfully lead to the high-eccentric orbit of the inner planet, typically larger than 0.99. For HJs in star clusters, the median value of the orbital period is around 3 days, which corresponds to 0.04 AU around a Solar-like star. If we set the final semi-major axis \citep[$a_{f}=2q_{\rm in}$,][]{2018MNRAS.479.5012O} after tidal circularization of a Jovian planet as 0.04 AU, the periastron of the inner planet $q_{\rm in}$ is about 0.02 au. According to Figure 1 or Equation 5 in \cite{2022MNRAS.509.5253W}, the typical tidal dissipation timescale of such a system is $\lesssim$ 100 Myr. Therefore, the typical formation timescale of HJs ($\tau_{\rm HJC}$) through flyby-induced high-e migration is $\lesssim$ 200 Myr under our parameter settings. Due to the uncertainty of the mass of outer companions, $\tau_{\rm HJC}$ may move backward to several hundreds of Myr. \textbf{In some cases, if we take account of chaotic or diffusive tides \citep[e.g.,][]{1995ApJ...450..722M,1995ApJ...450..732M}, planets may experience quicker tidal circularization than that under the scenario of equilibrium tides. I.e. the high-eccentricity migration process can be sped up in its early stages at high eccentricities, \cite{2018MNRAS.476..482V}. To sum up, HJs can form through flyby-induced high-e migration in open clusters within 200 Myr.} 

%$n_{*}$ can take a wide range of values, from 10$^{-1}$ stars pc$^{-3}$ in the nearby OB associations to 10$^{6}$ stars pc$^{-3}$ in the center of globular clusters. 
%which is consistent with our observation result.  the tidal circularization timescale should be extremely short($<<$ 100 Myr). Consequently, some of these young HJs ($\sim$ 100 Myr) may be probably with low eccentricity, while those formed a little bit later ($\sim$ 300 Myr) may have higher eccentricity on average. 

\cite{2018ARA&A..56..175D} provides that the occurrence rate of HJs is 0.5-1\%, which is ten times of estimation by \cite{2021MNRAS.508.3710R}. Interestingly, if we adopt the new stellar density value in \cite{2021ApJ...921...90P} (i.e. $n_{*}=10^4 \,\,\rm pc^{-3}$), the observed HJs can be successfully explained by flyby-induced high eccentricity migration in star clusters. This may indicate that the flyby-induced high-e migration is the dominant formation scenario of HJ in star clusters. 
%这一段重写一下很啰嗦。
\textbf{One intriguing system may be consistent with our scenario, i.e. Pr0211 b\&c. Pr0211, a member of Praesepe ($\sim$ 700 Myr) has two giant planets surrounding it. One is a Hot Jupiter within 3 days, i.e. Pr0211 b with near-circular orbit, and the other, Pr0211 c, is a cold Jupiter beyond 3,500 days with a very eccentric orbit ($e>0.6$, \cite{2016A&A...588A.118M}). \cite{2018A&A...610A..33P} demonstrates that a stellar fly-by scenario could shape the Pr0211 system successfully.} 

%, DS Tuc A b, TOI-1937 A b, and TOI-4145 A b. Specifically, P
%\begin{itemize}
%    \item Pr0211, a member of Praesepe (700 Myr) has two giant planets surrounding it. One is Hot Jupiter within 3 days, i.e. Pr0211 b with near-circular orbit, and the other (Pr0211 c) is a cold Jupiter beyond 3,500 days with a very eccentric orbit ($e>0.6$, \cite{2016A&A...588A.118M}). 
%    \item TOI-1937 A b is a Hot Jupiter within 1 day, whose host may be a potential member of NGC 2516 (150 Myr) \cite{2019AJ....158..122K}. The projection separation of the binary is about 1030 au \cite{2023ApJS..265....1Y}.  
%    \item TOI-4145 A b is a Hot Jupiter with a period of around 4 days, whose host may be a potential member Platais-3 (150 Myr) \cite{2019AJ....158..122K}.  The projection separation of the binary is about 356 au \cite{2023ApJS..265....1Y}.   
%    \item DS Tuc A b is a Saturn-sized planet in the 45 Myr Tucana–Horologium young moving group \cite{2019ApJ...880L..17N}. The binary separation is about 176 au. 
%\end{itemize}

\subsection{High-e migration of HJs around field stars beyond 1 Gyr}\label{subsec:4.2}
\begin{figure}
    \centering    
    \includegraphics[width=1\linewidth]{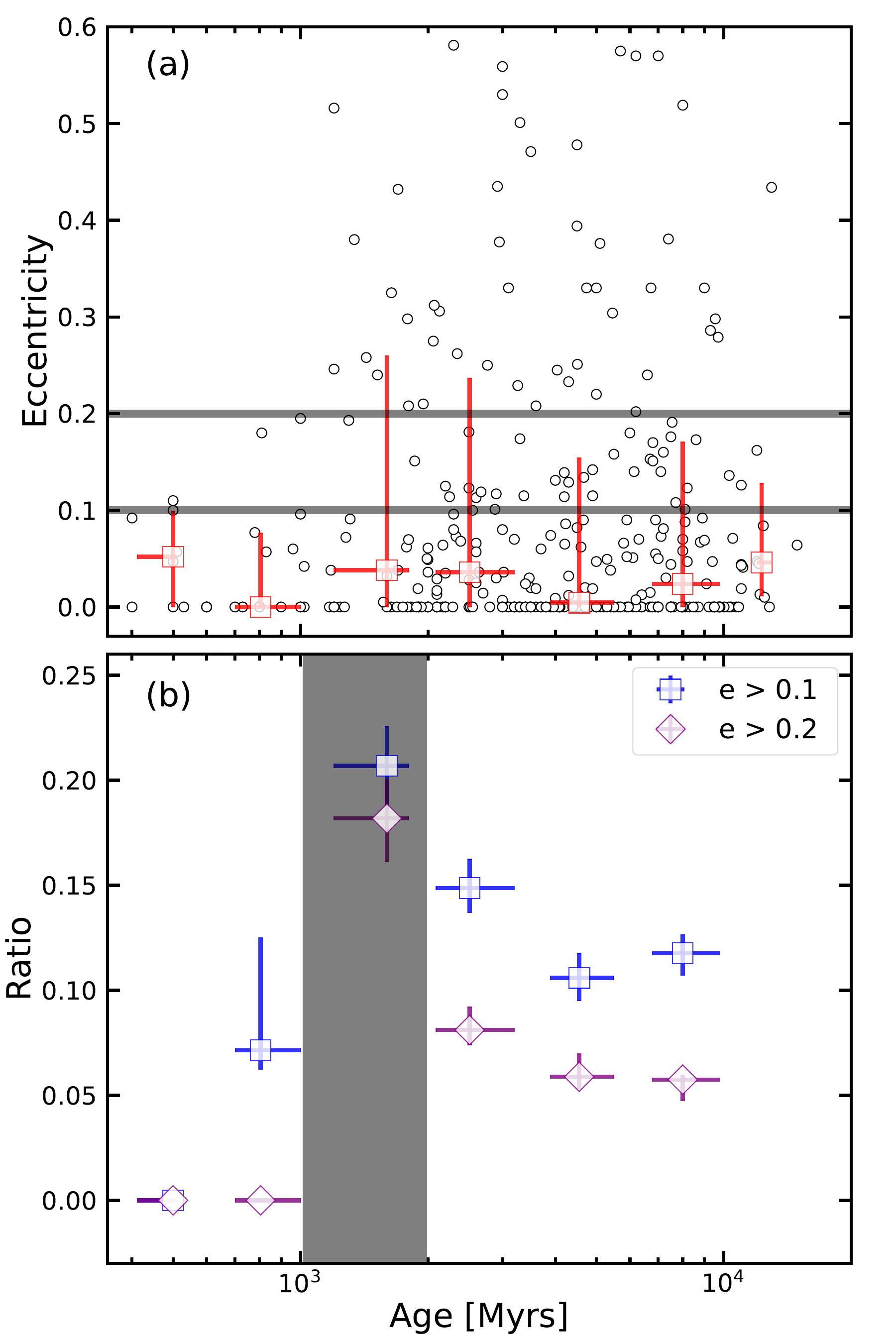}
    \caption{The eccentricity-age relation of HJs. We select 336 known HJs ($P<20$ days, $\sigma_{R_{p}}/R_{\rm p}<0.5$) with age and eccentricity measurements. Panel (a) shows the entire sample in the eccentricity-age diagram. Panel (b) represents the ratio of high-eccentricity HJs changing with age. The blue and purple dots are the ratio of HJs with $e>0.1$ and $e>0.2$ changing with age, respectively. The shadow region is over-plotted to emphasize the peak of the ratio. }
    \label{Figure:He_migration}
\end{figure}
%we we can see f$_{SubN}$ has a slight declination(i.e. 1 sigma difference of interval between the 7th and 9th red points) after 1 Gyr, and correspondingly, f$_{HJ}$ increases a little beyond 1 Gyrs and peaks at $\sim$ 1-3 Gyr(i.e. 1 sigma difference of interval between the 7th and 9th black points). Because f$_{SubS}$ is nearly unchanged during 1-3 Gyr, the anti-correlation between f$_{SubN}$ and f$_{HJ}$ is easy to understand. Yet the small peak of f$_{HJ}$ may be the physical results that are related to some dynamical mechanisms. If this small peak of f$_{HJ}$ is just a coincidence, there is probably no other coupling phenomenon. However, if it is truly related to some physical mechanisms, there should be some other observational results to be tested.
%One reasonable explanation is high-e migration. %\cite{2022MNRAS.509.5253W} mentioned that the tidal dissipation timescale is in a wide range, i.e. highly dependent on the initial semi-major axis and eccentricity. For HJs with periastron $\sim$ 0.05 au(i.e. 4 days around solar-like stars), its dissipation timescale will be older than the observable universe. Therefore, many HJs may still remain at high eccentricity after the dramatic high-e migration. 

In Figure \ref{Figure:relative}, we find a tiny bump of proportion of Jovian planets $f_{\rm J}$ around 2 Gyr, which is anti-correlated with $f_{\rm SubN}$, i.e. a small dip of $f_{\rm SubN}$ around 2 Gyr. 

High-e migration can explain the anti-correlation. During the inward migration of the Jovian planet, the inner planets can be ejected from the system due to the planet-planet interaction \citep{2015ApJ...808...14M}. I.e. HJs after high-e migration are usually lonely \citep{2012PNAS..109.7982S,2021AJ....162..263H}. Thus, the proportion of smaller planets, e.g. Sub-Neptunes and Super-Earths, have a declination.  

Due to tidal dissipation during the high-e migration, the eccentricity of the HJs will decrease with time. Therefore, if the tiny bump in the proportion of Jovian planets $f_{\rm J}$ around 2 Gyr is attributed to the HJs that form through high-e migration, we may also see a small bump in the eccentricity-age diagram around 2 Gyr.

To test this conjecture, in the following, we select 336 HJs both having eccentricity and age measurements from the NASA exoplanet archive. We constrain the period of the HJs to be shorter than 20 days and the relative uncertainty of planet radii is no more than 50\%. Figure \ref{Figure:He_migration} shows the eccentricity-age distribution of the 336 HJs. In Panel (a) red dots show the median eccentricity of each age bin changing with age. The error bar is calculated according to the 16 and 84 percentiles of the eccentricities of each age bin. Because the majority of the HJs have low eccentricity, the median eccentricity of each age bin is nearly a zero constant, i.e. unchanging with age. 

However, red dots have larger error bars in the older region. It seems that the number of HJs with high eccentricity increases beyond 1 Gyr. Therefore, we calculate the relative ratio of High-eccentricity HJs in each age bin, as shown in panel (b) of Figure \ref{Figure:He_migration}. Similar to the previous analysis, this ratio is calculated with the assumption of the Gaussian distribution of age and eccentricity of each HJ.  We find that both the ratio of HJs with $e>0.1$ and $e>0.2$ rapidly increase beyond 1 Gyrs, i.e. the difference between the two data points before and after 1 Gyr is 2 $\sigma$ and 12 $\sigma$, respectively. After reaching the maximum of around 2 Gyr, the ratio of HJs with high-eccentricity declines with age due to the tidal dissipation as expected. Therefore, the eccentricity evolution also supports the high-e migration for these HJs older than 1 Gyr.

Because HJs around field stars are the dominant population beyond 1 Gyr, as shown in Figure \ref{Figure:sizeage}, we can conclude that the bump of $f_{\rm J}$ around 2 Gyr is likely due to the formation of HJs around field stars through high-e migration. Since the timescale of 2 Gyr is ten times longer than the formation timescale of HJs through flyby-induced high-e migration in star clusters ($\tau_{\rm HJC} \lesssim $ 200 Myr). We explain the large differences via the different environments of stellar density. The HJs form around field stars after 1 Gyr, which escape the cluster environments at the early stage ($<100$ Myr). Because the less dense dynamical environments lead to longer flyby timescale, the trigger time of high-e migration is probably much longer than that in cluster environments.

\subsection{Formation of Hot Neptune desert around 100 Myr}\label{subsec:4.3}
Several previous works find the hot Neptune desert in planetary mass-period distribution and radius-period distribution, i.e. \cite{2011ApJ...727L..44S,2013ApJ...763...12B,2016A&A...589A..75M}. In Figure \ref{Figure:relative}, we find that the proportion of the Sub-Jupiters within 20 days, $f_{\rm SubJ}$, rapidly declines around 100 Myr, then remains at the low value. The declination is correlated to the hot Neptune desert and may indicate the formation timescale of such desert.

\begin{figure}
    \centering        
    \includegraphics[width=1\linewidth]{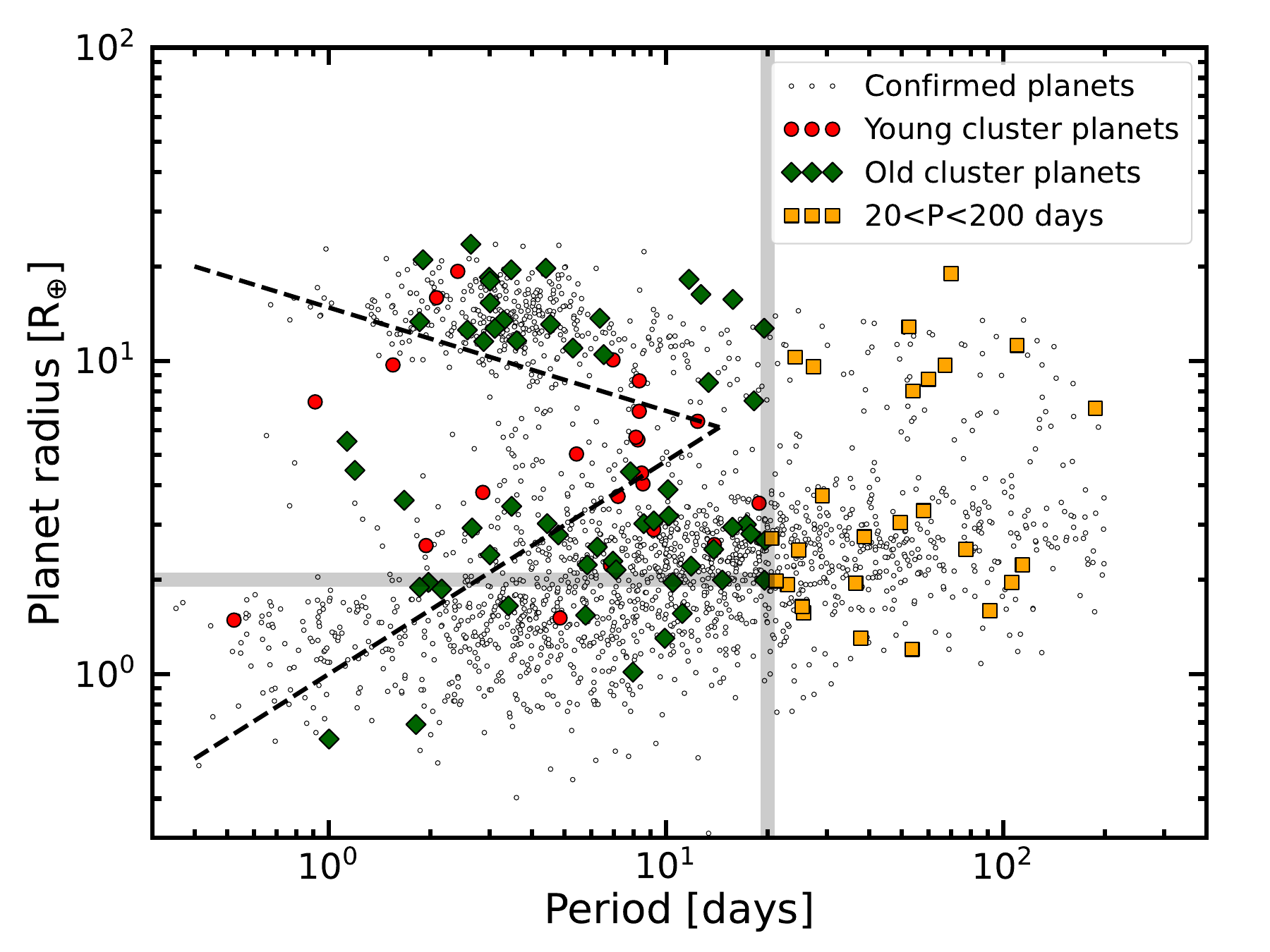}
    \caption{The scatter plot of planets in planets radius-period plane. Red dots are hot and young planets in star clusters($P<20$ days, age $<$ 100 Myr). While black dots are hot and old planets in star clusters ($P<20$ days, age $>$ 100 Myr). Orange squares are planets in clusters with periods beyond 20 days. Two black dashed lines are boundaries of the Hot Neptune desert described in \cite{2016A&A...589A..75M}. The gray line of P=20 days and the gray line of $R_{\rm p}$=2 R$_{\oplus}$ are over-plotted as reference.}
    \label{fig:hot-Neptune}
\end{figure}

However, the Sub-Jupiters classified via radius and period independently, are not in the hot-Neptune desert exactly. According to \cite{2016A&A...589A..75M}, the borders of the desert are period-dependent. I.e. at large radii, the planet's radius decreases with increasing period, while at small radii, the radius increases with increasing period (the dashed lines in Figure \ref{fig:hot-Neptune}). 
Using the same region with \cite{2016A&A...589A..75M}, we compare the time-dependent ratio of the number of planets inside and outside the hot Neptune desert, to constrain the formation timescale of the hot Neptune desert. 

In Figure \ref{fig:hot-Neptune}, we show 107 planets/candidates in star clusters and 1991 other confirmed planets around field stars in the radius--period plane. We divide the planets/candidates in star clusters within 20 days into two groups. One is planets younger than 100 Myr (red circles). The other is planets older than 100 Myr (green diamonds). 

We calculate the ratio of the number of planets inside and outside the hot Neptune desert, i.e. $N_{\rm in}/N_{\rm out}$, for both the younger and older groups. Then, we use the Monte Carlo simulation to get a distribution of $N_{\rm in}/N_{\rm out}$ under the assumption of the Gaussian distribution of planet radius, period, and age. The error of $N_{\rm in}/N_{\rm out}$ is adopted according to the 16 and 84 percentiles of the ratio distribution. We find that the younger group have much higher $N_{\rm in}/N_{\rm out}$ (0.80$^{+0.20}_{-0.19}$) than older groups (0.19$^{+0.03}_{-0.03}$) in nearly 3.0 $\sigma$ confidence level. If we add a planet radius cut for detection completeness, i.e. $R_{\rm p} >2 \rm R_{\oplus}$ (horizontal shadow line), the result is similar. I.e. the younger group  has higher $N_{\rm in}/N_{\rm out}$ (0.67$^{+0.17}_{-0.13}$) than older group (0.18$^{+0.03}_{-0.01}$) in 3.67 $\sigma$ confidence level. Alternatively, if we use 300 Myr to distinguish the younger and older group, the difference of $N_{\rm in}/N_{\rm out}$ between the young and old group will decrease, which is smaller than the 100-Myr case. I.e. the difference between the median value of the young group's $N_{\rm in}/N_{\rm out}$ (0.48$^{+0.07}_{-0.05}$), and old group's $N_{\rm in}/N_{\rm out}$ (0.31$^{+0.04}_{-0.04}$). Therefore, we can conclude that the rapid declination of $f_{SubJ}$ around 100 Myr corresponds to the formation of the hot-Neptune desert around 100 Myr. 

\cite{2018MNRAS.479.5012O} explains two boundaries of the hot-Neptune desert in a combination of photoevaporation and high-e migration. For the lower boundary, photoevaporation functions effectively in the first several hundred of million years and will trigger the atmospheric mass loss of Neptune-sized planets, which is consistent with the formation timescale we obtained. If the high-e migration sculpts the lower boundary, the formation timescale will be much larger than 100 Myr, because the hot-Neptunes usually experience longer tidal circularization ($\gtrsim$ 1 Gyr). We prefer that photoevaporation sculpts the lower boundary of the hot-Neptune desert around 100 Myr.  

%需要考虑场星中desert的影响
For the upper boundary, photoevaporation seems not a suitable explanation. Because several works show that the massive planets(M$_{p}>$ 0.5 M$_{J}$) can resist photoevaporation even at extremely short periods \citep[e.g.][]{2012MNRAS.425.2931O,2015ApJ...808..173T,2016ApJ...816...34O}. Therefore, photoevaporation will predict a lower upper boundary(Figure 4 in \cite{2018MNRAS.479.5012O}). However, in some specific cases, the rapid declination of the radius of giant planets may explain the upper boundary of the Hot-Neptune desert. E.g.  \cite{2023ApJ...945L..36T} developed a model including radius inflation, photoevaporative mass loss, and Roche lobe overflow, which can trigger a runaway mass loss of a puff hot Saturn around 400 Myr. Because the desert already existed around 100 Myr, we do not consider such a mechanism as the dominant.  
%？, such as Roche-lobe overflow triggered by vigorous photoevaporation of close-in giant planets (\cite{2014ApJ...783...54K}). However, the Roche-lobe overflow happens several Gyr typically in \cite{2014ApJ...783...54K} which is much longer than our timescale constraints, 100 Myr. Recently, \cite{2023ApJ...945L..36T} developed a model that includes radius inflation, photoevaporative mass loss, and Roche lobe overflow. In their model, puff hot Saturn with a density smaller than 0.1 g cm$^{-3}$ can experience a runaway mass loss at the age of 400 Myr. Because the timescale is model-dependent, for hotter planets around an active star, the runaway mass loss may happen earlier. Therefore, such Roche-lobe overflow could be another explanation for the upper boundary of the hot Neptune desert.

High-e migration can deliver Jovian planets from the outer region to the inner orbit. Then, the subsequent decay due to stellar tides will further sculpt the upper boundary. More specifically, both the tidal circularization timescale (planetary tides) and tidal decay timescale (stellar tides) decrease with increasing planetary radius or mass. As described in section \ref{subsec:4.1}, the typical formation timescale of HJs in clusters is $\lesssim$ 200 Myr, which is consistent with the formation timescale of the Hot-Neptune desert. Therefore, flyby-induced high-e migration could sculpt the upper boundary of the hot-Neptune desert in clusters.

The formation timescale of the Hot-Neptune desert in star clusters is around 100 Myr. When it comes to field stars, the formation timescale of the Hot-Neptune desert may move backward to several Gyr because of the relatively slow high-e migration for HJs around field stars. However, due to the limited sample of young planets around field stars, we do not find any evidence. 

Because the scenario of flyby-induced high-e migration can not only explain the increment of $f_{\rm J}$ around 100 Myr, but may also sculpt the upper boundary of the Hot-Neptune desert around 100 Myr. Therefore, we prefer to the scenario of flyby-induced high-e migration, i.e. a combination of photoevaporation and flyby-induced high-e migration could sculpt the hot-Neptune desert around 100 Myr.

%In Figure \ref{Figure:relative}, we found that the proportion of sub-Saturn-sized planets f$_{\rm SubS}$ declines rapidly around 100 Myr, then remain at the low value. This result indicates that the formation timescale of the hot-Neptune desert may be around 100 Myr. Photoevaporation functions effectively in the first 100 Myr because the X-ray fluxes of the majority of solar-like stars will undergo a saturated state in the first 100 Myr, then decline rapidly.  In other words, the rapid declination of f$_{SubS}$ can be attributed to a combination of photoevaporation and flyby-induced high-e migration in star clusters around 100 Myr. 

% Both photoevaporation and high-e migration can sculpt the lower boundary in the radius-period plane. However, low-mass planets can only be produced by high-e migration followed by tidal circularization in a very narrow range of parameter space. Therefore, we tend to use photoevaporation to explain the lower boundary. For the upper boundary, . While high-e migration works for massive M$_{p}>$ 0.2 M$_{J}$ and can produce a similar upper boundary.  

\subsection{The paucity of young WJs between 100 and 1000 Myr}\label{subsec:4.4}
\begin{figure}
    \centering       
    \includegraphics[width=1\linewidth]{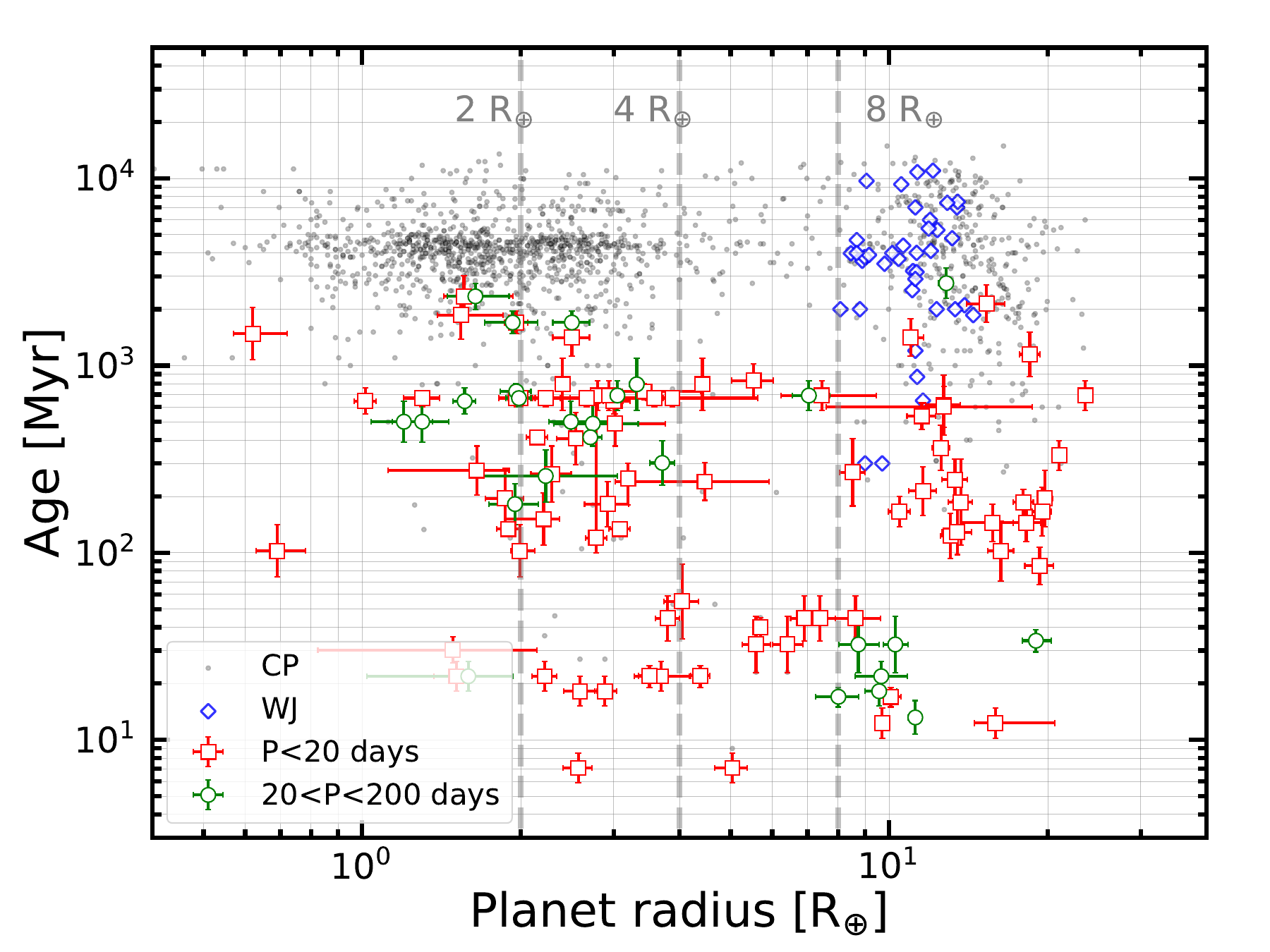}
    \caption{The planetary radius -- age distribution of 107 planets/candidates (within 200 days, red squares and green circles) in clusters and 1991 confirmed planets (within 200 days, gray dots) around field stars. Here, we also show the 39 warm Jupiters around field stars with age measurements from the NASA exoplanet archive ($20<P<200$ days, blue diamonds). }
    \label{Figure:WJ}
\end{figure}

%在前面的论述中 
\textbf{In the analysis above, we find that high-eccentricity migration may play an important role in the formation of HJs and the Hot-Neptune Desert in star clusters. During the high-e migration, some progenitors of HJs (i.e. warm-Jupiters and cold-Jupiters) may migrate inward and become HJs. Consequently, the number of warm-Jupiters and cold-Jupiters may decline after $\sim$100 Myr. To test the conjecture, we include warm planets in clusters within 200 days. There are 7 young Warm-Jupiters (hereafter WJ, $20 < P < 200$ days, age $<$ 100 Myr) in clusters as shown in Figure \ref{Figure:WJ} (green circles). However, there are no WJs in clusters with ages between 100 and 1000 Myr. This may be another hint of the flyby-induced high-e migration in star clusters.} 

\textbf{Note there are dozens of WJs with age measurements from the NASA exoplanet archive ($20<P<200$ days, 8$\rm R_{\oplus}<R_{\rm p}<2.5 \rm {R_{\rm J}}$, blue diamonds in Figure \ref{Figure:WJ}). The majority of these WJs around field stars are older than 1 Gyr. The absence of WJs between 100--1000 Myr may indicate that WJs may hardly survive in cluster environments, i.e. such relatively dense environments prefer the formation of HJs.} 

\textbf{One should be cautious of these preliminary results, the absence of WJs between 100 Myr and 1000 Myr may be due to the observation bias. For instance, TESS prefers to discover HJs instead of WJs, because of the relatively short time span. Apart from that, the inhomogeneous age measurements from the NASA exoplanet archive may also be considered as potential observational bias. Therefore, these preliminary but intriguing results need more observation data to check.}

%快速引出我们的结论。
\section{Discussion} \label{sec:5}
\textbf{In section \ref{sec:4}, we have explained our statistical results through flyby-induced high-e migration and photoevaporation. However, there are still some factors that we have not discussed. In the following, we discuss the influence of the number of stars (section \ref{subsec:5.1}) and some caveats of eccentricity measurement for young planets (section \ref{subsec:5.2}).}

\subsection{The evolution of planet radius considering the number of stars}\label{subsec:5.1}
\textbf{Note, the number of stars varies in different clusters, which may influence the occurrence of the planet to some extent. For example, \cite{2023Univ....9..192F} uses TESS data to constrain the HJ occurrence rate in associations. However, it's out of our concern in this paper. Therefore, we do not use the planet occurrence rate to describe the evolution of planet radius, which will be discussed further in our next paper. Here, we simply use the fraction of planets with different sizes in each cluster to describe the influence of star number of the parental clusters, i.e. }

\begin{equation}
    F(R_{p},t) = \Sigma \frac{N_{p}}{N_{s}}
\end{equation}

\begin{figure}
    \centering    
    \includegraphics[width=1\linewidth]{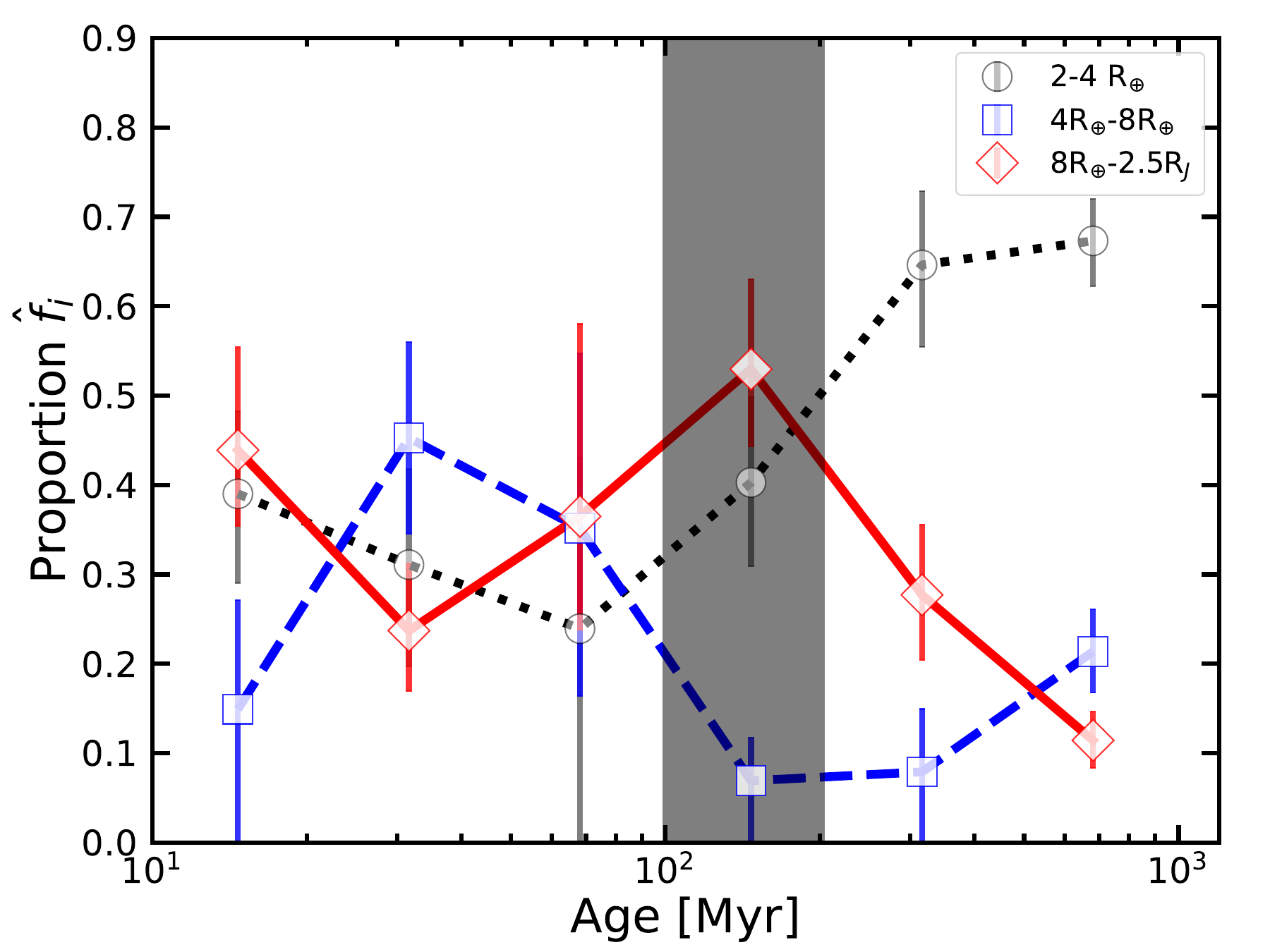}
    \caption{The time-dependent relation of the proportions of the fraction of planets of different sizes in different star clusters. Different colors show planets of different sizes. The shadow region around 100 Myr is over-plotted to emphasize the typical timescale. }
    \label{Figure:Np2Ns}
\end{figure}

\textbf{where $N_{p}$ is the number of planets in one cluster and $N_{s}$ is the number of cluster member stars. Similar to the calculation of $f_{i}$, i.e. proportions of planets with different sizes and ages, we use Monte Carlo simulation to derive the proportion of fraction of planets with different sizes and ages, i.e. $\hat{f}_{i}$.}

\begin{equation}
    \hat{f}_{\rm i} = \frac{F_{\rm i}}{F_{\rm SubN}+F_{\rm SubJ}+F_{\rm J}},
\end{equation}
\textbf{where $F_{i}$ is the fraction of planets in star clusters with different sizes and ages, i.e. $F_{\rm SubN}, F_{\rm SubJ}$ or $F_{\rm J}$ corresponding to Sub-Neptunes, Sub-Jupiters, and Jovian planets in section \ref{subsec:3.1}, respectively. Figure \ref{Figure:Np2Ns} shows the results of $\hat{f}_{i}$, which is similar to Figure \ref{Figure:cluster}. $\hat{f}_{\rm J}$ (red diamonds) increases before 100 Myr and then declines after 200 Myrs, i.e. a peak occurs between 100 and 200 Myr. $\hat{f}_{\rm SubJ}$ (blue squares) declines around 100 Myr. For Sub-Neptunes $\hat{f}_{\rm SubN}$ (gray circles) shows a clear increase after 100 Myr. }

%\subsubsection{Short time span of TESS} \ref{sec:dis timespan}

\subsection{Eccentricity measurement of young planets} \label{subsec:5.2}
%关于hot Jupiter eccentricity的讨论。
\textbf{Planet eccentricity is more challenging to measure for young planets because of magnetic activity. In several cases (e.g. TOI-837 b in open cluster IC 2602 \cite{2020AJ....160..239B}) circular orbits are simply adopted to reduce the number of parameters in the fitting.} 

\textbf{Additionally, planets with high impact parameter $b$ are hard to derive the accurate eccentricity distribution, because of the degeneracy with other orbital parameters, e.g. b>0.9 for TOI-837 b \cite{2020AJ....160..239B}.}

\textbf{In our sample, we exclude giant planets with orbital periods longer than 20 days. These planets may include some progenitors of Hot Jupiters with highly eccentric orbits, which are going through tidal circularization and shrinking their orbit. Only including the progenitors of HJs will make the total sample more complete. However, due to the small time span of TESS, the progenitors of HJs are not detected completely. This problem with be much less sensitive for Kepler. Using high eccentricity progenitors of HJs from Kepler data, \cite{2023AJ....165...82J} finds that high-eccentricity migration is probably the dominant formation channel of HJs, yet fails to account for all of the HJs. Therefore, we emphasize that excluding of progenitors of Hot Jupiters will have some influence, yet will not change our qualitative results of eccentricity.}

\section{Conclusion} \label{sec:6}
Planets in young star clusters could help us understand the planet's formation and evolution because of the accurate age estimation. In section \ref{sec:2}, we collect the largest catalog of 73 planets and 84 candidates in star clusters by cross-matching with K2020 and planets/candidates from the NASA exoplanet archive. We validate the age estimation of 70 planets/candidates in star clusters, obtain more convinced ages of three host stars via either literature or the new gyrochronological relation, and exclude eight planetary systems with no robust age estimations. 
 
%Planets in young star clusters could shed light on planet formation and evolution since star clusters can provide accurate age estimation. Although many surveys and projects are focusing on planets around young stars, the total number of planets in star clusters is still too small for statistical study. Thanks to the accurate astrometry data of Gaia DR2 and Gaia DR3, many newly identified open clusters(or comoving groups) can enlarge membership in open clusters, as the planets in stars clusters. In this paper, we use the star catalog from \cite{2020AJ....160..279K} to cross-match known planet catalogs including confirmed planets and planet candidates. After carefully removing false positives, we collect and obtain the largest catalog of planets in star clusters, i.e. 70 confirmed planets and 79 planet candidates. 
In section \ref{sec:3}, we use this catalog to study the planet radius -- age relation. The main statistical results are as follows:
\begin{itemize}
    \item The proportion of Jovian planets $f_{\rm J}$ increases around 100 Myr and reaches a maximum between 100 Myr and 200 Myr, which is mainly attributed to the HJs in star clusters. The bump of $f_{\rm J}$ around 2 Gyr is attributed to the HJs around field stars. 
    \item The proportion of Sub-Jupiters $f_{\rm SubJ}$ declines rapidly around 100 Myr, then remains at a low value. The declination of $f_{\rm SubJ}$ is mainly attributed to the hot Sub-Jupiters in star clusters.
\end{itemize}

After discussing several possible scenarios to explain the results, we give two constraints on the hot giant exoplanet formation timescales in section \ref{sec:4}:
\begin{itemize}
    \item HJs likely form through flyby-induced high-e migration in star clusters within 200 Myr. 
    \item A combination of photoevaporation and flyby-induced high-e migration in star clusters can sculpt the hot-Neptune desert around 100 Myr.   
\end{itemize}

We find that flyby-induced high-e migration may be the dominant formation channel of HJs in star clusters. As described in section \ref{subsec:4.1}, those HJs in star clusters will accompany an outer companion, which is an effective angular momentum transmitter during a close flyby event. Therefore, we hope to discover some outer companions beyond these HJs with the Radial Velocity observations from ground-based telescopes and the astrometric data from the future data release of Gaia. 

Different from HJs in star clusters, HJs around field stars may have much longer formation timescale ($\sim$ 2 Gyr), which can be attributed to the different dynamical environments (section \ref{subsec:4.2}). 

Note in this paper, we mainly focus on the ZLK mechanism to excite the high eccentricity of the inner Jovian planet. Actually, other mechanisms such as planet-planet scattering can trigger the high-eccentric orbit too. \cite{2020ApJ...905..136W} demonstrates that a very small fraction of HJs can form from the fly-induced planet-planet scattering channel, i.e. ZLK mechanism may be the dominant scenario of eccentricity excitation. However, we could not exclude the possibility of planet-planet scattering. One way to distinguish these two mechanisms definitely is the stellar obliquity, the angle between a planet's orbital axis and its host star's spin axis. ZLK mechanism predicts a bimodal stellar obliquity distribution, concentrated at 40$^{\circ}$ and 140$^{\circ}$ \citep[e.g.][]{2007ApJ...669.1298F}. While the planet-planet scattering after a convergent disk migration predicts a concentration of stellar obliquity around 90$^{\circ}$ \citep{2011ApJ...742...72N}. 

Additionally, a hint from the absence of WJs around field stars between 100 Myr and 1000 Myr also supports the scenario of flyby-induced high-e migration (section \ref{subsec:4.1}). However, this absence may be due to the observation bias \ref{subsec:4.4}. For instance, TESS prefers to discover HJs instead of WJs, because of the relatively short time span. 

In the future, with the extended mission of TESS, the Earth 2.0 mission \citep[ET2.0,][]{2022arXiv220606693G}, the Chinese Space Station Telescope \citep[CSST,][]{2011SSPMA..41.1441Z,2019ApJ...883..203G}, and PLATO \citep{2014ExA....38..249R}, we hope to detect more young planets both in star clusters and around field stars. The subsequent astrometry data from Gaia and the follow-up Radial-Velocity observation (including the Rossiter-McLaughlin effect) from ground-based telescopes can also provide more information about warm planets and even outer companions. A larger sample of planets in clusters will benefit us to test different formation scenarios of HJs, as well as hot-Neptune deserts. 

\acknowledgments
\textbf{We thank the anonymous referee and Prof. Dr. Bo Ma for helpful recommendations to improve the paper.} This work is supported by the National Natural Science Foundation of China (grant Nos. 11973028, 11933001, 1803012, 12150009) and the National Key R\&D Program of China (2019YFA0706601). We also acknowledge the science research grants from the China Manned Space Project with No. CMS-CSST-2021-B12 and CMS-CSST-2021-B09, as well as Civil Aerospace Technology Research Project (D050105).


\vspace{5mm}
\software{astropy \citep{2013A&A...558A..33A}, 
          matplotlib \citep{Hunter:2007},
          pandas \citep{mckinney-proc-scipy-2010},
          Lightcurve \citep{2018ascl.soft12013L}.
          }


\newpage
\appendix
\section{SNR-age relation} \label{appendix:a}%?这个部分我在想是否需要放到前面去讨论，逻辑上是没啥问题。/data refinement
%The other bias is that younger stars have higher stellar activities(i.e. larger noise) which results in lower detection completeness. Here, we do not focus on the first bias, because for young planets that are embedded in gas disks, no whether matter they are larger or small, we can hardly detect them through the transiting method. We focus on the second bias. 

In young clusters, stars are more active and may hide the transiting events, especially for small planets. Thus, the detection of small planets in young clusters is uncompleted. To study the selection effects, we try to derive the empirical relations between SNR and stellar age for planets of different sizes. Then, we can select suitable criteria of planet radius to cut our samples.  
 
The calculation of SNR of a transiting planet is calculated as follows: 

\begin{equation}
\rm {SNR} = \frac{\delta n^{0.5}}{\sigma_{*}\left(t_{\rm dur}\right)}
\end{equation}
where $\delta=(R_{p}/R_{*})$ is the transit depth of the star and $n$ is the transit number. $\sigma_{*}$ is the stellar photometric noise. The transit duration time $t_{\rm dur}$ is given by:

\begin{equation}
t_{\rm dur} = \frac{PR_{*}\sqrt{1-e^{2}}}{\pi a}
\end{equation}
where $P$ is the planet's orbital period and $a$ is the semi-major axis. In the calculation of SNR, we assume that host stars are solar-like (i.e. $M_{*} = 1 \rm M_{\odot}$ and $R_{*} = 1 \rm R_{\odot}$ ) and set $P$=20 days (because most of our select planets in star clusters are within 20 days). The same as the assumption of the \cite{2015ApJ...798..112M} to assume that the stellar noise change with time as a simple power law distribution:

\begin{equation}
\sigma_{*}\left(t\right) = \sigma_{\rm LC}\left(\frac{t}{t_{\rm LC}}\right)^{ind_{\rm CDPP}},
\end{equation}
where we normalize the noise ($\sigma_{\rm LC}$) at the period in the long cadence, 1765.5 s ($t_{\rm LC}$). $ind_{\rm CDPP}$ is the power law index. We use Combined Differential Photometric Precision \citep[CDPP,][]{2012PASP..124.1279C} from Kepler DR25, which characterizes the noise level in Kepler lightcurves. Then, using the stellar kinematic age from \cite{2021AJ....162..100C}, we can obtain $ind_{\rm CDPP}$--age relation and SNR--age relation.

For a rough calculation, we assume that the stars observed by Kepler and TESS are similar, i.e. the stellar noise evolution is similar. The major difference in SNR of planets of different sizes lies in the observation time $t_{obs}$ which determines the transit numbers. Here, the observation time of the Kepler stars is $\sim 1450$ days and TESS is roughly two observation sectors, i.e. $\sim 54$ days. 

\begin{figure}
    \renewcommand{\thefigure}{A1}
    \centering
    \includegraphics[width=0.95\linewidth]{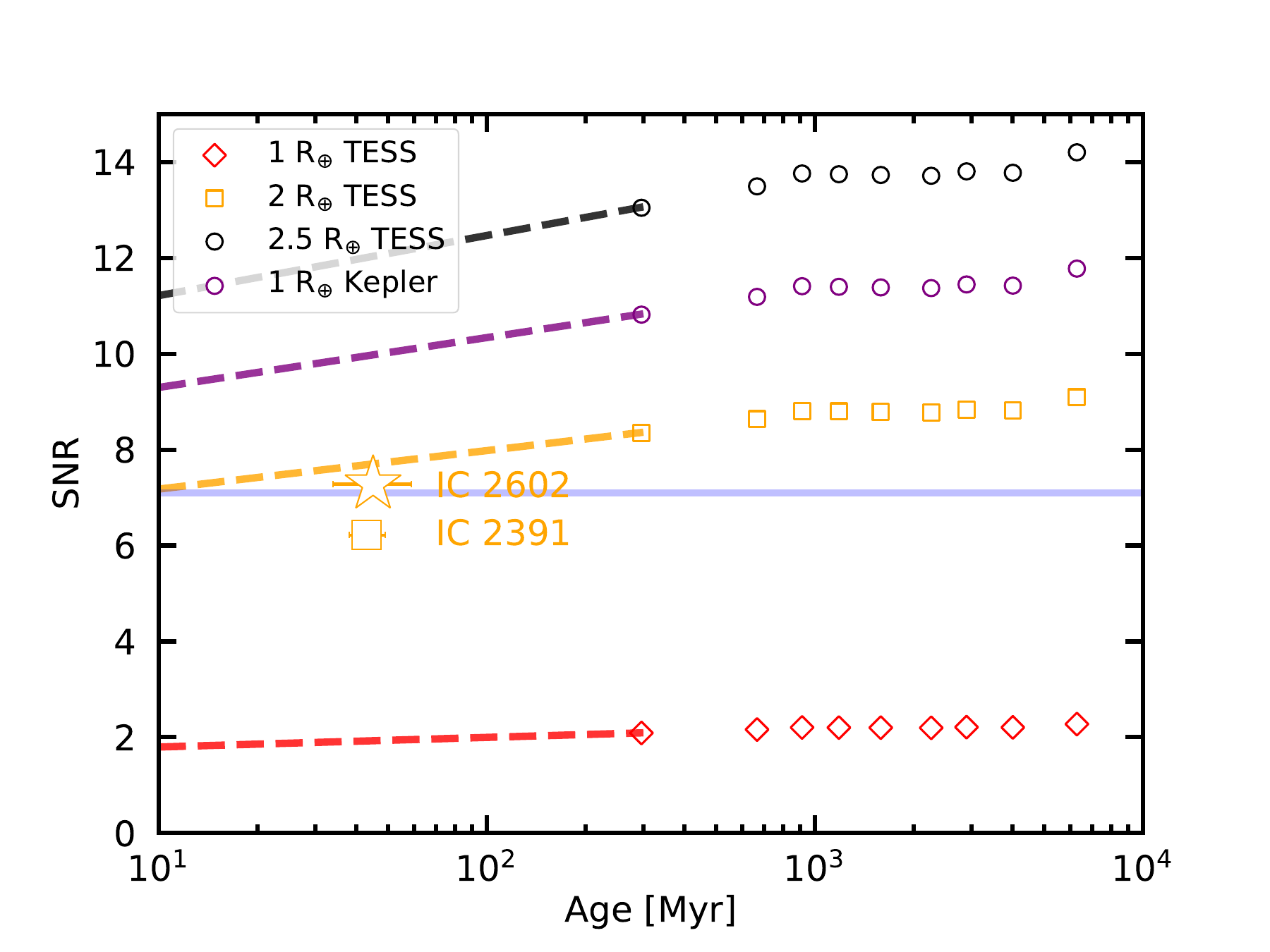}
    \caption{The SNR-age relation of planets of different sizes. Red, orange, and black hollow dots and dashed lines are SNR-age relations of planets of 1, 2, and 2.5 R$_{\oplus}$ for TESS. The purple presents the SNR-age relation of planets of 1 R$_{\oplus}$ for Kepler. \textbf{Young open cluster IC 2602 ($\sim$ 40 Myr, orange star) and IC 2391 ($\sim$ 40 Myr, orange square) are plotted as examples.} The blue horizontal line (SNR=7.1) is over-plotted as a reference.}
    \label{Figure:SNR}
\end{figure}

Figure \ref{Figure:SNR} shows the calculated SNR of planets changing with age. Red, orange, and black hollow dots and dashed lines present the results of planets of different sizes (i.e. 1 R$_{\oplus}$, 2 R$_{\oplus}$, and 2.5 R$_{\oplus}$) for TESS. The purple hollow dots and dashed line show the result of planets of 1 R$_{\oplus}$ for Kepler. Since the data from \cite{2021AJ....162..100C} does not provide the CDPP of stars younger than 300 Myr, we simply extend the relationship to 10 Myr through a log-linear exploration. The blue horizontal line is the SNR of 7.1 above which we consider TESS or Kepler can detect planets. \textbf{To validate our empirical SNR-age relation between 10 and 300 Myr, i.e. the log-linear extrapolation, we calculate the CDPPs of the solar-type stars in two young open clusters, i.e. IC 2602 ($\sim$ 40 Myr, orange star in Figure \ref{Figure:SNR}) and IC 2391 ($\sim$ 40 Myr, orange square). Here, the stellar noise CDPP is calculated through the simpler “sgCDPP proxy algorithm” discussed by \cite{2011ApJS..197....6G}. In Figure \ref{Figure:SNR}, we show the median SNRs of detecting 2 R$_{\oplus}$ sized planets around solar-type stars in young open clusters. The average observation time span for these two clusters is $\sim$ 100 days. However, to compare the log-near exploration in Figure \ref{Figure:SNR}, we assume that the observation time span for these two clusters is $\sim$ 54 days. In other words, the true SNRs of IC 2602 and IC 2391 are higher than the threshold 7.1. Thus, it is reasonable to constrain the planets above 2 R$_{\oplus}$ for the estimate of completeness of transit detection.} Therefore, in section \ref{subsec:2.2}, we focus on the planets with a radius larger than 2 R$_{\oplus}$ to exclude the uncompleted detection due to the stellar noise of young stars. 

\section{Poisson Error} \label{appendix:b}
\begin{figure}
    \renewcommand{\thefigure}{B1}
    \centering
    \includegraphics[width=0.95\linewidth]{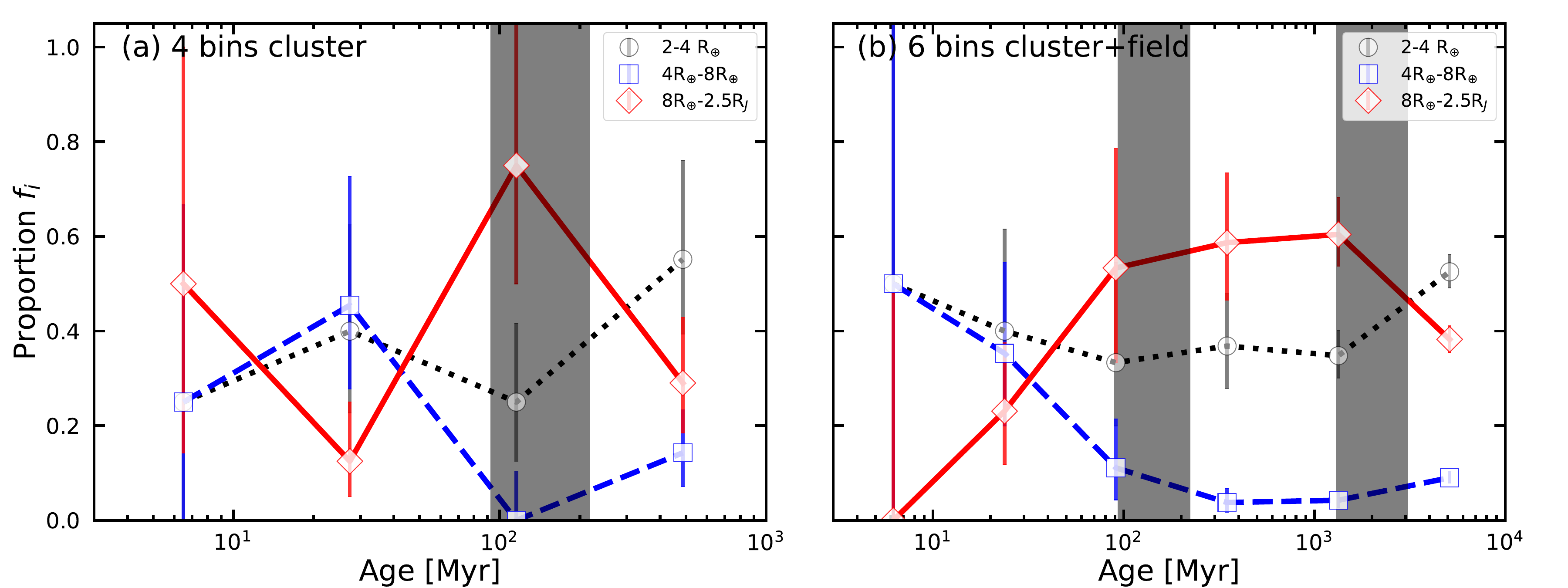}
    \caption{The time-dependent relation of the proportions of planets of different sizes. Here, the error is adopted via Poisson distribution. Panel (a) corresponds to Figure \ref{Figure:cluster} and panel (b) corresponds to Figure \ref{Figure:relative}. }
    \label{Figure:poisson}
\end{figure}

The ``standard'' confidence interval for a Poisson parameter
Figure \ref{Figure:poisson} shows the time-dependent relation of the proportions of planets of different sizes, adopted Poisson error. The error bar, i.e. standard confidence interval related to a Poisson parameter, is calculated through the chi-square distribution. Panel (a) includes planets/candidates in star clusters, consistent with the results of Figure \ref{Figure:cluster}, although with more significant uncertainties. Panel (b) includes planets/candidates in star clusters and around field stars, consistent with the results of Figure \ref{Figure:relative}. I.e. the $f_J$ increases before 100 Myr and then decreases around 1-2 Gyr. 

\section{The radius evolution of confirmed planets} \label{appendix:c}
\begin{figure}
    \renewcommand{\thefigure}{C1}
    \centering
    \includegraphics[width=0.95\linewidth]{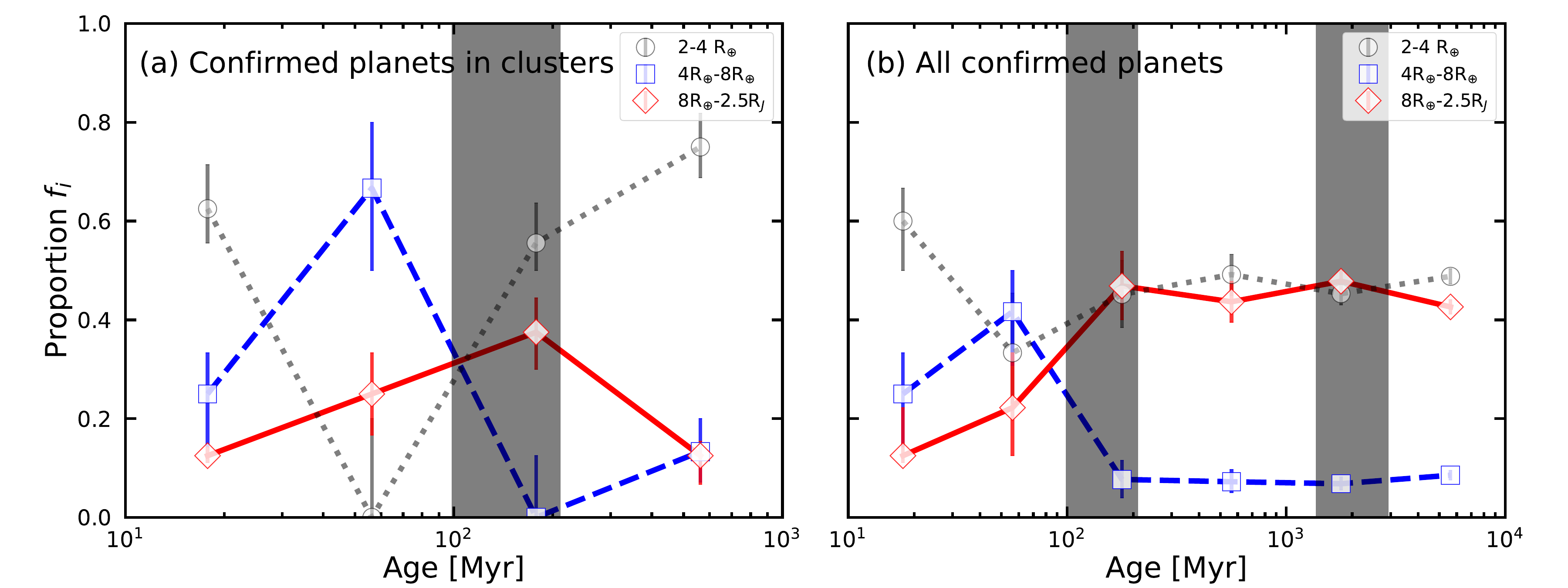}
    \caption{The time-dependent relation of the proportions of planets of different sizes. The error is adopted via MC simulation as Figure \ref{Figure:cluster}. Panel (a) uses confirmed planets in star clusters and panel (b) uses all confirmed planets.}
    \label{Figure:confirmed}
\end{figure}
\newpage
%\begin{figure*}
%    \centering
%    \includegraphics[width=1\linewidth]{age_rp_diagram.pdf}
%    \caption{The planetary radius -- age distribution of 103 planets and planet candidates in star clusters(Table 2). The gray dots are confirmed transit planets whose host stars have age measurements. Different colors and markers show the planets discovered by different facilities.}
%    \label{Figure:sizeage}
%\end{figure*}

%\begin{figure*}
%    \centering    
%    \includegraphics[width=1\linewidth]{age_ratio_last_.pdf}
%    \caption{The relative proportions of planets of different sizes changing with age. Panel (a) and (b) uses 35 planets and 50 planet candidates in star clusters. We add a cut of stellar effective temperature in panel (c) and an additional period cut in panel (d). }
%    \label{Figure:relative}
%\end{figure*}

\bibliography{reference}{}
\bibliographystyle{aasjournal}


\end{document}